\documentclass[letterpaper, nontitlepage, 12pt, twocolumn]{aastex63}
\usepackage{enumerate}
\usepackage{amsthm}
\usepackage{calligra} 
\usepackage{braket}
\usepackage{mathtools}
\usepackage{tikz}
\usepackage{cancel}
\usepackage{booktabs}
\usepackage{topcapt}
\usepackage{comment}
\graphicspath{{images/}}
\theoremstyle{definition}

\allowdisplaybreaks[1]
\DeclareMathAlphabet{\mathcalligra}{T1}{calligra}{m}{n} 
\DeclareFontShape{T1}{calligra}{m}{n}{<->s*[2.2]callig15}{}   

\newcommand{\R}{\mathbb{R}}

\newcommand*{\colorboxed}{}
\def\colorboxed#1#{%
  \colorboxedAux{#1}%
}
\newcommand*{\colorboxedAux}[3]{%
  \begingroup
    \colorlet{cb@saved}{.}%
    \color#1{#2}%
    \boxed{%
      \color{cb@saved}%
      #3%
    }%
  \endgroup
}

\newcommand{\HRB}{HR~2562~B}

\begin{document}
\title{Testing the Interaction Between a Substellar Companion and a Debris Disk in the HR~2562 System}


\author[0000-0003-2122-6714]{Stella Yimiao Zhang}
\affiliation{Center for Astrophysics and Space Sciences, University of California, San Diego, La Jolla, CA 92093, USA}
\author[0000-0002-5092-6464]{Gaspard Duch\^{e}ne}
\affiliation{Astronomy Department, University of California, Berkeley; Berkeley, CA 94720, USA}
\affiliation{Univ. Grenoble Alpes/CNRS, IPAG, F-38000 Grenoble, France}
\author[0000-0002-4918-0247]{Robert J. De Rosa}
\affiliation{European Southern Observatory, Alonso de C\'{o}rdova 3107, Vitacura, Santiago, Chile}
\author[0000-0003-4142-9842]{Megan Ansdell}
\affiliation{NASA Headquarters, 300 E Street SW, Washington, DC 20546, USA} 
\author[0000-0002-9936-6285]{Quinn Konopacky}
\affiliation{Center for Astrophysics and Space Sciences, University of California, San Diego, La Jolla, CA 92093, USA}
\author[0000-0002-0792-3719]{Thomas Esposito} 
\affiliation{Astronomy Department, University of California, Berkeley; Berkeley, CA 94720, USA}
\affiliation{SETI Institute, Carl Sagan Center, 189 Bernardo Av, Suite 200, Mountain View CA 94043, USA}
\author[0000-0002-6246-2310]{Eugene Chiang}
\affiliation{Department of Astronomy, UC Berkeley, Berkeley CA, 94720, USA}
\author[0000-0002-7670-670X]{Malena Rice}
\affiliation{Department of Astronomy, Yale University, New Haven, CT 06511, USA}
\author{Brenda Matthews}
\affiliation{Herzberg Astronomy and Astrophysics, National Research Council of Canada, 5071 West Saanich Rd., Victoria, BC V9E 2E7, Canada}
\author[0000-0002-6221-5360]{Paul Kalas} 
\affiliation{Astronomy Department, University of California, Berkeley; Berkeley, CA 94720, USA}
\affiliation{SETI Institute, Carl Sagan Center, 189 Bernardo Av, Suite 200, Mountain View CA 94043,USA}
\affiliation{Institute of Astrophysics, FORTH, GR-71110 Heraklion, Greece}
\author[0000-0003-1212-7538]{Bruce Macintosh} 
\affiliation{Kavli Institute for Particle Astrophysics and Cosmology, Department of Physics, Stanford University, Stanford, CA, 94305, USA} 
\author[0000-0001-7016-7277]{Franck Marchis} 
\affiliation{SETI Institute, Carl Sagan Center, 189 Bernardo Av, Suite 200, Mountain View CA 94043, USA} 
\author[0000-0003-3050-8203]{Stan Metchev} 
\affiliation{Department of Physics \& Astronomy, Institute for Earth and Space Exploration, The University of Western Ontario, London, ON N6A 3K7, Canada.} 
\author{Jenny Patience} 
\affiliation{School of Earth and Space Exploration, Arizona State University, PO Box 871404, Tempe, AZ 85287, USA} 
\author[0000-0003-0029-0258]{Julien Rameau} 
\affiliation{Univ. Grenoble Alpes/CNRS, IPAG, F-38000 Grenoble, France}
\author[0000-0002-4479-8291]{Kimberly Ward-Duong} 
\affiliation{Smith College, 10 Elm Street, Northampton, MA 01063, USA} 
\author[0000-0002-9977-8255]{Schuyler Wolff} 
\affiliation{Steward Observatory, University of Arizona, Tucson, AZ 85721, USA} 

\author[0000-0002-0176-8973]{Michael P. Fitzgerald}
\affiliation{Department of Physics \& Astronomy, University of California, Los Angeles, CA 90095, USA}
\author[0000-0002-5407-2806]{Vanessa P. Bailey}
\affiliation{Jet Propulsion Laboratory, California Institute of Technology, 4800 Oak Grove Dr., Pasadena, CA 91109, USA }
\author[0000-0002-7129-3002]{Travis S. Barman}
\affiliation{Lunar and Planetary Lab, University of Arizona, Tucson, AZ 85721, USA}
\author{Joanna Bulger}
\affiliation{Institute for Astronomy, University of Hawai’i, 2680 Woodlawn Drive, Honolulu, HI 96822, USA}
\author[0000-0002-8382-0447]{Christine H. Chen}
\affiliation{Space Telescope Science Institute, 3700 San Martin Drive, Baltimore MD 21218 USA}
\author[0000-0001-6305-7272]{Jeffrey K. Chilcotte}
\affiliation{Department of Physics and Astronomy, University of Notre Dame, 225 Nieuwland Science Hall, Notre Dame, IN, 46556, USA }
\author{Tara Cotten}
\affiliation{Department of Physics and Astronomy, University of Georgia, Athens, GA 30602, USA}
\author[0000-0001-5485-4675]{Ren\'e Doyon}
\affiliation{Institut de Recherche sur les Exoplan\`etes, D\'epartement de physique, Universit\'e de Montr\'eal, Montr\'eal, QC H3C 3J7, Canada}
\author[0000-0002-7821-0695]{Katherine B. Follette}
\affiliation{Physics and Astronomy Department, Amherst College, 25 East Drive, Amherst, MA 01002, USA}
\author[0000-0003-3978-9195]{Benjamin L. Gerard}
\affiliation{University of Victoria, Department of Physics and Astronomy, 3800 Finnerty Rd, Victoria, BC V8P 5C2, Canada}
\affiliation{National Research Council of Canada Herzberg, 5071 West Saanich Rd, Victoria, BC, V9E 2E7, Canada }
\author[0000-0002-4144-5116]{Stephen Goodsell}
\affiliation{Department of Physics, Durham University, Stockton Road, Durham DH1, UK}
\affiliation{Gemini Observatory, Casilla 603, La Serena, Chile}
\author{James R. Graham}
\affiliation{Astronomy Department, University of California, Berkeley; Berkeley, CA 94720, USA}
\author[0000-0002-7162-8036]{Alexandra Z. Greenbaum}
\affiliation{IPAC, Mail Code 100-22, Caltech, 1200 E. California Blvd., Pasadena, CA 91125, USA}
\author[0000-0003-3726-5494]{Pascale Hibon}
\affiliation{European Southern Observatory, Alonso de C\'{o}rdova 3107, Vitacura, Santiago, Chile}
\author{Li-Wei Hung}
\affiliation{Natural Sounds and Night Skies Division, National Park Service, Fort Collins, CO 80525, USA}
\author[0000-0003-3715-8138]{Patrick Ingraham}
\affiliation{Vera C. Rubin Observatory, 950 N Cherry Ave, Tucson AZ, 85719, USA}
\author{J\'er\^ome Maire}
\affiliation{Center for Astrophysics and Space Sciences, University of California, San Diego, La Jolla, CA 92093, USA}
\author[0000-0002-5251-2943]{Mark S. Marley}
\affiliation{Space Science Division, NASA Ames Research Center, Mail Stop 245-3, Moffett Field CA 94035, USA }
\author[0000-0002-4164-4182]{Christian Marois}
\affiliation{Herzberg Astronomy and Astrophysics, National Research Council of Canada, 5071 West Saanich Rd., Victoria, BC V9E 2E7, Canada}
\affiliation{University of Victoria, Department of Physics and Astronomy, 3800 Finnerty Rd, Victoria, BC V8P 5C2, Canada}
\author[0000-0001-6205-9233]{Maxwell A. Millar-Blanchaer}
\affiliation{Department of Physics, University of California, Santa Barbara, Santa Barbara, CA, USA}
\author[0000-0001-6975-9056]{Eric L. Nielsen}
\affiliation{Department of Astronomy, New Mexico State University, 1320 Frenger Mall, Las Cruces, NM 88003-8001, USA}
\author[0000-0001-7130-7681]{Rebecca Oppenheimer}
\affiliation{American Museum of Natural History, Department of Astrophysics, Central Park West at 79th Street, New York, NY 10024, USA}
\author{David W. Palmer}
\affiliation{Lawrence Livermore National Laboratory, 7000 East Ave, Livermore, CA, 94550, USA}
\author[0000-0002-3191-8151]{Marshall D. Perrin}
\affiliation{Space Telescope Science Institute, 3700 San Martin Drive, Baltimore, MD 21218, USA}
\author{Lisa A. Poyneer}
\affiliation{Lawrence Livermore National Laboratory, 7000 East Ave, Livermore, CA, 94550, USA}
\author{Laurent Pueyo}
\affiliation{Space Telescope Science Institute, 3700 San Martin Drive, Baltimore, MD 21218, USA}
\author[0000-0002-9246-5467]{Abhijith Rajan}
\affiliation{Space Telescope Science Institute, 3700 San Martin Drive, Baltimore, MD 21218, USA}
\author[0000-0002-9667-2244]{Fredrik T. Rantakyr\"o}
\affiliation{Gemini Observatory, Casilla 603, La Serena, Chile}
\author[0000-0003-2233-4821]{Jean-Baptiste Ruffio}
\affiliation{Department of Astronomy, California Institute of Technology, Pasadena, CA 91125, USA}
\author[0000-0002-8711-7206]{Dmitry Savransky}
\affiliation{Sibley School of Mechanical and Aerospace Engineering, Cornell University, Ithaca, NY 14853, USA}
\author[0000-0002-6294-5937]{Adam C. Schneider}
\affiliation{US Naval Observatory, Flagstaff Station, P.O. Box 1149, Flagstaff, AZ 86002, USA.}
\affiliation{Department of Physics and Astronomy, George Mason University, MS3F3, 4400 University Drive, Fairfax, VA 22030, USA}
\author[0000-0003-1251-4124]{Anand Sivaramakrishnan}
\affiliation{Space Telescope Science Institute, 3700 San Martin Drive, Baltimore, MD 21218, USA}
\author[0000-0002-5815-7372]{Inseok Song}
\affiliation{Department of Physics and Astronomy, University of Georgia, Athens, GA 30602, USA}
\author[0000-0003-2753-2819]{Remi Soummer}
\affiliation{Space Telescope Science Institute, 3700 San Martin Drive, Baltimore, MD 21218, USA}
\author[0000-0002-9121-3436]{Sandrine Thomas}
\affiliation{Vera C. Rubin Observatory, 950 N Cherry Ave, Tucson AZ, 85719, USA}
\author[0000-0003-0774-6502]{Jason J. Wang}
\affiliation{Department of Astronomy, California Institute of Technology, Pasadena, CA 91125, USA}
\affiliation{Center for Interdisciplinary Exploration and Research in Astrophysics (CIERA) and Department of Physics and Astronomy, Northwestern University, Evanston, IL 60208, USA}
\author[0000-0003-4483-5037]{Sloane J. Wiktorowicz}
\affiliation{ Remote Sensing Department, The Aerospace Corporation, 2310 E. El Segundo Blvd., El Segundo, CA 90245}


\begin{abstract}
The HR\,2562 system is a rare case where a brown dwarf companion resides in a cleared inner hole of a debris disk, offering invaluable opportunities to study the dynamical interaction between a substellar companion and a dusty disk.  We present the first ALMA observation of the system as well as the continued GPI monitoring of the companion's orbit with 6 new epochs from 2016 to 2018. We update the orbital fit and, in combination with absolute astrometry from GAIA, place a 3$\sigma$ upper limit of 18.5 $M_J$ on the companion's mass. To interpret the ALMA observations, we used radiative transfer modeling to determine the disk properties. We find that the disk is well resolved and nearly edge on. While the misalignment angle between the disk and the orbit is weakly constrained due to the short orbital arc available, the data strongly support a (near) coplanar geometry for the system. Furthermore, we find that the models that describe the ALMA data best have an inner radius that is close to the companion's semi-major axis. Including a posteriori knowledge of the system's SED further narrows the constraints on the disk's inner radius and place it at a location that is in reasonable agreement with, possibly interior to, predictions from existing dynamical models of disk truncation by an interior substellar companion. 
HR\,2562 has the potential over the next few years to become a new testbed for dynamical interaction between a debris disk and a substellar companion.
\end{abstract}


\section{Introduction}

Debris disks are the gas poor disk structures surrounding stars as the outcome of the star formation process.
The presence of a debris disk suggests that there needs to be larger bodies colliding and grinding down the dust grains in order to sustain the disk. Furthermore, observable features like gaps and warps in disk morphology can be the results of the dynamical interaction between a substellar companion and the disk itself. These features provide another pathway for us to find and study the properties of these  potential planets residing near the disk. The most readily available example of this would be our solar system where substructures within the Asteroid Belt and the Kuiper Belt were created from the resonance between the planet orbits and smaller objects \citep[e.g.,][]{Kirkwood}. 
 A great amount of information can be extracted from studying the dynamical interactions that would create such features in morphology in disk structures, even in the case where the companions are not directly detected. Sophisticated dynamical models have been developed to constrain properties of the companion from the properties of the disk and the system.

However, it is in fact difficult to find appropriate systems to test these models. Many of the currently discovered debris disks often do not have the required resolution to study the disk structures, and in those resolved disks, the detected companions are often found too far away from the disruption site to be solely responsible. Out of the systems where a companion is detected close to the irregular structure of the disk, there often isn't convincing evidence for disk-companion interaction. As of today, only a few systems were discovered to be candidates for disk-companion interaction. The first one of this kind is $\beta$ Pictoris, a system with an almost perfectly edge on debris disk with a vertical warp at 85 au, and a planet $\beta$ Pictoris b that is consistent with this warp \citep{Smith:1987ea,Augereau2006,Lagrange2019,Lagrange2010}. It is believed that the inner warp of the disk is a result of the dynamical interaction between the planet $\beta$ pic b and the disk, but studying this interaction is difficult due to the radial density profile being model-dependent \citep{Dawson2011,Augereau2006}. Only a handful of similar systems were discovered in the decades following the discovery of $\beta$ Pictoris. Those systems all have potential dynamically interacting planets/brown dwarf candidates residing within the disks’ rings, but all of them lack determining pieces of evidence due to observational limitations or other possible scenarios for the disk structure formation \citep{Wilner2018,Wang2021,Barcucci2019,Chauvin2018,Su2017}.

So far, only two systems have directly imaged brown dwarf companions orbiting inside the debris ring: HD~206893 \citep{Nederlander2021,Marino2020} and HR~2562. In the former, it is unclear how close the companion is to the disk and whether it is responsible for truncating it. In this paper, we focus on analyzing the potential for dynamical interaction in HR~2562. HR~2562 is an F5V star with a mass of 1.3$M_{\odot}$, located at 34pc away from the Sun \citep{Gray2006,Casagrande2011,vanLeeuwen:2007dc}. Its debris disk was first imaged by Herschel \citep{Moor2006}. In 2016, Konopacky and the GPI team were able to directly image and obtain data for the orbit of the substellar companion, HR~2562 B, residing within the inner hole of the disk with GPIES. The orbit of HR~2562 B was further monitored by VLT/SPHERE for 10 months from 2016-2017, providing support for a coplanar geometry \citep{Maire:2018ch}, and its spectral type was characterized as T2-T3 by IFS and IRDIS of VLT/SPHERE \citep{Mesa2018}. By comparison to evolutionary models, \citet{Konopacky:2016dk} estimated a mass of 30$\pm$15\,$M_{\text{jup}}$ for the companion, with the uncertainty dominated by the poorly constrained age of the system.
However, concrete evidence of brown dwarf-disk interaction would require a determination of the inner radius of the disk, which the initial characterization of the disk was unable to constrain due to limited angular resolution \citep{Moor:2015kf}. This led to an uninformative upper limit of $\sim 0.24M_{\odot}$ based solely on dynamical arguments \citep{Konopacky:2016dk}.
Continued monitoring of the orbit as well as better resolved observation of the disk are required to 
better characterize the disk, derive evidence for dynamical interaction and further constrain the properties of the companion.

In this paper, we present the new Atacama Large Millimeter/submillimeter Array (ALMA) observation of the HR~2562 system (section \ref{sec:alma-obs}) as well as the updated GPI observation of the companion (section \ref{sec:GPI Observation}).  We present the reduced ALMA image in section \ref{sec:ALMA image} and discuss the companion orbit fit, the coplanar scenario as well as the dynamical mass limit in section \ref{sec:orbit result}. The Monte Carlo Markov Chain (MCMC) fit for the ALMA image and the analysis are detailed in section \ref{sec:disk_modeling}, and further discussion on the system geometry, SED selection of the best fitting models and the dynamical interaction are presented in section \ref{sec:analyis}. We summarize our results and conclude in section \ref{sec:conclusion}. 

\section{Observation}
\subsection{ALMA Observation}
\label{sec:alma-obs}
We observed the HR~2562 debris disk with the Atacama Large Millimeter Array (ALMA, project 2016.1.00880.S, PI: G. Duch\^ene) on May 15th, 2018. To maximize sensitivity for faint dust thermal emission in the system, we observed the target in Band 7 for a total on-source integration time of 37\,min. Observing conditions were good with 0.8\,mm of precipitable water vapor at zenith. As a compromise between resolving the {\sl Herschel}-estimated inner radius of the disk and minimizing spatial filtering on large scales, we selected configuration C40-1 with the 12m-array, with 44 antennas providing baselines ranging from 14 to 313\,m and an angular resolution of about 1\farcs1. As this is a pure continuum observation, we used four 2\,GHz bandwidth channels, centered at 336.5, 338.5, 348.5 and 350.5\,GHz. Observations in all four bands are ultimately combined in a single continuum map at an effective frequency of 343.5\,GHz (870\,$\mu$m).

\begin{figure*}
\includegraphics[width=\textwidth]{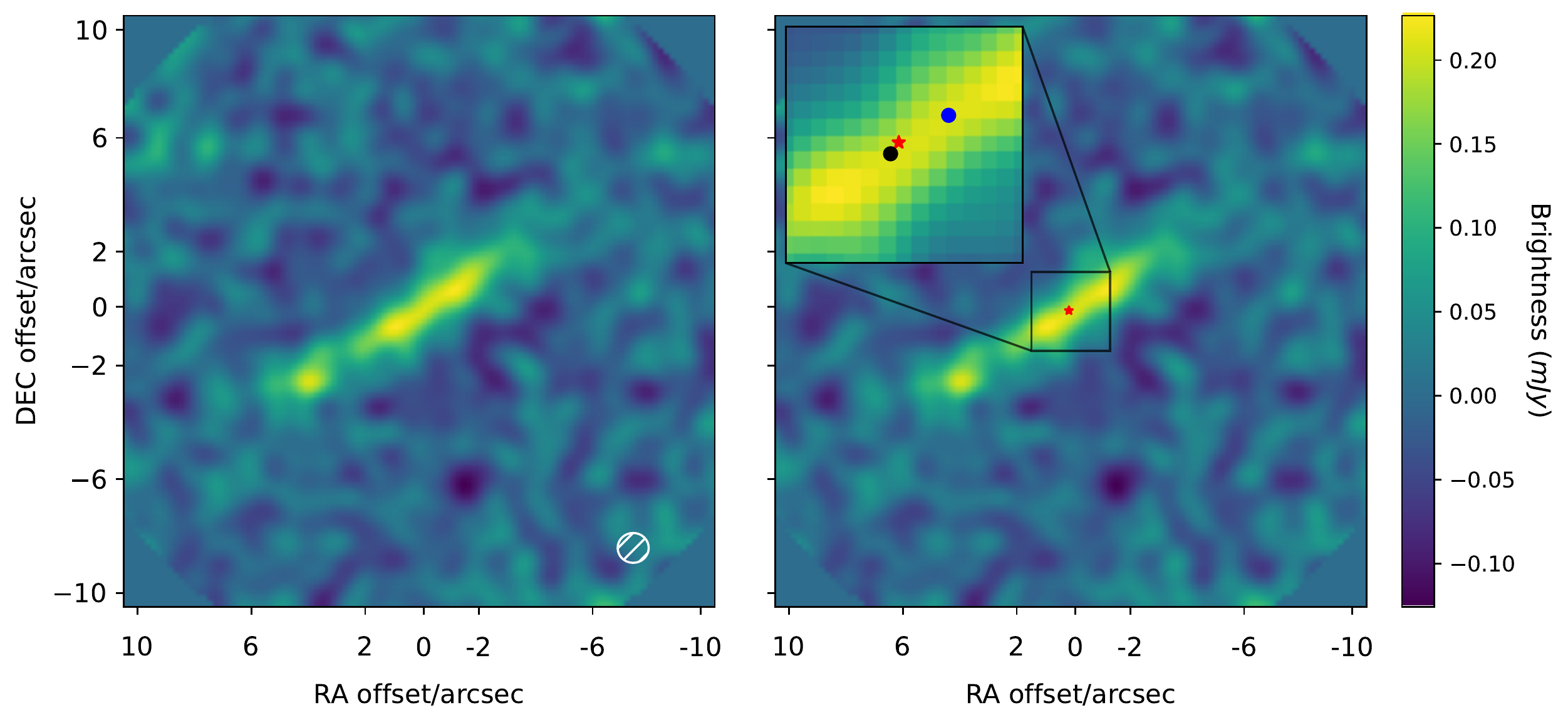}

\caption{Left panel: ALMA data of the HR~2562 system at 870 micron. The (0,0) of the images correspond to RA= 102.5\degr, DEC=-60.2\degr; Right panel:a zoomed in region of the center of the disk is shown on the top left corner. The black circle represents the center of the disk determined by 2D Gaussian fit; red star represents the Gaia DR3 coordinates of the star HR~2562; blue circle represents the orbit predicted location of the companion HR~2562~B on Jan 31st, 2018. The disk center and the star are well aligned, and the companion is close to the star as expected.\label{fig:Data}}
\end{figure*}

The data were processed 
with standard routines from the Common Astronomy Software Applications \citep[CASA,][]{McMullin2007}, version 5.3.0-143. Specifically, we applied phase, bandpass and flux calibrations using the provided calibrators. 
To produce the final continuum map, we used the task {\sl tclean} with Briggs weighting with robust=0.5.
This results in a beam size of 1\farcs10$\times$1\farcs06 at a position angle of -132\fdg3 and a rms of 0.0323 mJy/beam. 
The left panel of Figure\,\ref{fig:Data} shows the final ALMA map generated with an 0\farcs2 pixel scale. 

\subsection{GPI Observation}
\label{sec:GPI Observation}

\begin{deluxetable*}{ccccccccc}
\tablecaption{HR 2562 Gemini/GPI observing log and associated KLIP parameters\label{tbl:gpi-log}}
\tablehead{\colhead{UT Date} & \colhead{Filter} & \colhead{$N_{\rm exp}$} & \colhead{$t_{\rm int}$} & \colhead{$\Delta$PA\tablenotemark{a}} & 
\colhead{$\lambda_{\rm min}$--$\lambda_{\rm max}$\tablenotemark{b}} & \colhead{$n_{\lambda}$\tablenotemark{c}} & \colhead{$m$\tablenotemark{d}} & \colhead{$n_{\rm KL}$\tablenotemark{e}}\\
 & & & (sec.) & (\degr) & ($\micron$) & & (px) & \\}
\startdata
2016 Jan 25\tablenotemark{$\dagger$} & $H$ & 33 & $59.6$ & 19.4 & 1.514--1.778 & 35 & 3 & 5\\
2016 Jan 28\tablenotemark{$\dagger$} & $K_1$ & 21 & $59.6$ & 10.6 &  1.947--2.173 & 26 & 3 & 5\\
2016 Jan 28\tablenotemark{$\dagger$} & $K_2$ & 20 & $59.6$ & 10.5 &  2.119--2.226 & 15 & 3 & 5\\
2016 Feb 25\tablenotemark{$\dagger$} & $K_2$ & 43 & $59.6$ & 25.7 &  2.116--2.224 & 15 & 3 & 5\\
2016 Feb 28\tablenotemark{$\dagger$} & $J$ & 53 & $59.6$ & 26.6 &  1.137--1.330 & 35 & 3 & 5\\
2016 Dec 17 & $K_1$ & 85 & $59.6$ & 40.7 & 1.940--2.173  & 26 & 4 & 5\\
2017 Feb 13 & $H$ & 19 & $59.6$ & 12.3 & .507--1.778 & 35 & 3 & 5\\
2017 Nov 29 & $H$ & 50 & $59.6$ & 30.1 &  1.511--1.773 & 35 & 3 & 5\\
2018 Jan 31 & $K_2$ & 54 & $88.7$ & 32.9 & 2.113--2.228 & 15 & 4 & 5\\
2018 Mar 10 & $Y$ & 66 & $59.6$ & 42.0 &  0.957--1.132 & 35 & 3 & 5\\
2018 Nov 19 & $H$ & 24 & $59.6$ & 12.5 &  1.509--1.779 & 35 & 3 & 5\\
\enddata
\tablenotetext{\dagger}{Re-reduction of observations presented in \citet{Konopacky:2016dk}.}
\tablenotetext{a}{Total field rotation over the duration of the observing sequence.}
\tablenotetext{b}{Full spectral range used in the data reduction.}
\tablenotetext{c}{Number of independent spectral channels.}
\tablenotetext{d}{Minimum rotation-induced displacement for inclusion in the PSF subtraction process.}
\tablenotetext{e}{Number of KLIP modes used in the PSF subtraction process.}
\end{deluxetable*}

\subsubsection{Observations \& Initial reduction}

The Gemini Planet Imager \citep[GPI,][]{Macintosh:2014js} is an instrument equipped with a high-order adaptive optics (AO) system \citep{Poyneer:2014ki,Poyneer:2016kb}, an apodized Lyot coronagraph \citep{Soummer:2011eq}, and both a dispersing and a Wollaston prism for spectroscopic and polarimetric observations. The instrument was designed to achieve high contrast at small angular separations, providing sensitivity to substellar companions and circumstellar material around nearby, bright stars. HR~2562 was observed with GPI on 11 separate epochs between 2016 and 2018 under program IDs GS-2015B-Q-501 and GS-2017B-Q-501. The first four epochs were originally analyzed and published in \cite{Konopacky:2016dk} but are re-reduced and analyzed here to ensure consistency. The observing strategy was similar for each dataset. The target was observed with GPI's coronagraphic mode with the specific coronagraph optimized for the near-infrared filter being used. After the coronagraph, a lenslet array and dispersing prism was used to disperse the light at each point within the $2\farcs8\times2\farcs8$ field into a low-resolution ($R\sim$35--80) spectrum. The resulting dispersed field was imaged with GPI's integral field spectrograph \citep[IFS,][]{Chilcote:2012hd,Larkin:2014ek}. In each dataset the star itself was used as the AO guide star. The observations were timed to be taken close to the meridian passage of the star over the observatory to maximize field rotation, $\Delta$PA, for angular differential imaging \citep[ADI,][]{Marois:2006df}. Observations of an argon lamp were taken during the target acquisition to measure the instrument flexure induced by the changing gravity vector as the telescope changes position. Standard dark and wavelength calibration frames were taken during the daytime as a part of the observatory's calibration plan. A summary of the observations are given in Table~\ref{tbl:gpi-log}.

The raw and associated calibration data were reduced through two separate processes. The raw data were reduced using the GPI Data Reduction Pipeline \citep[DRP,][]{Perrin:2014jh} v1.5.0 revision \texttt{cafd46a}. This version of the pipeline resolved several issues identified with the calculation of the average parallactic angle during an exposure \citep{DeRosa2020:kd}. Although the magnitude of this correction was small for the datasets presented in \cite{Konopacky:2016dk}, between 0\fdg09--0\fdg12 (see Figure 17 in \citealp{DeRosa2020:kd}), we took the opportunity to reprocess these data along with the new epochs presented in this work. The data were reduced using the procedure outlined in \citep{DeRosa:2020gy}. Briefly, the raw images were dark subtracted and bad pixels were identified using a combination of a static bad pixel map and outlier rejection and replaced. The microspectra within each image were extracted to create a three-dimensional $(x,y,\lambda)$ data cube containing the low-resolution spectra at each point within the field of view. Additional bad pixel identification was performed using outlier rejection. Distortion and anamorphism was corrected using a static distortion map applied to each slice within the data cube. The position and brightness of the star within each slice of each reduced data cube was estimated by measuring the four satellite spots generated by a wire diffraction grid within the pupil plane \citep{Sivaramakrishnan:2006cf}. The calibration data used to reduce the science data were not affected by the pipeline changes described in \cite{DeRosa2020:kd}. We used the dark frames and wavelength calibrations generated by the GPIES Data Cruncher \citep{Wang:2018co} created using v1.4.0 of the pipeline to reduce the science data rather than reprocessing them with the updated pipeline.

\subsubsection{PSF subtraction and astrometry}

\begin{figure}
   \centering
   \includegraphics[width=0.5\textwidth]{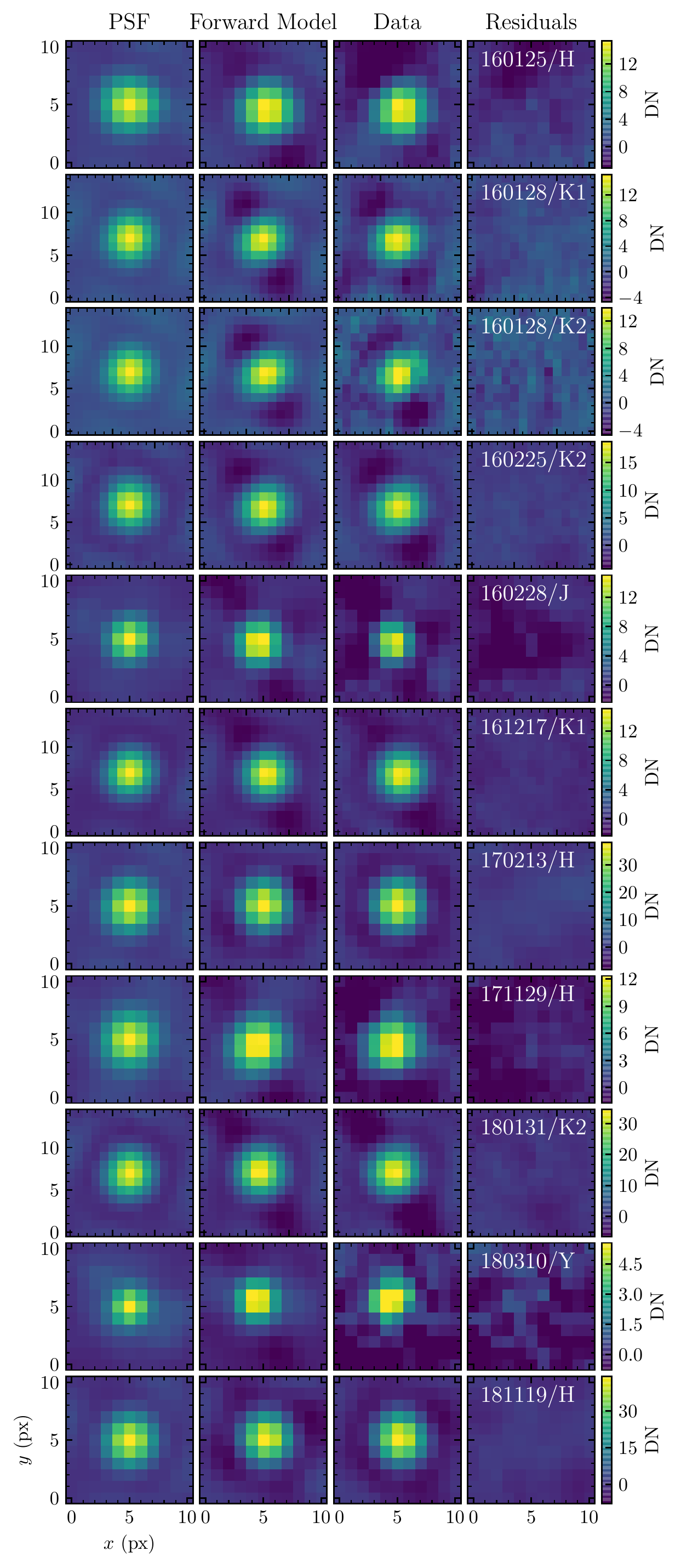} 
   \caption{The GPI PSF estimated from the four satellite spots (left column), BKA forward model (second column), the GPI images of HR 2562 B (third column), and residuals (fourth column) for each observation. The color scales are different for each dataset and are denoted by the color bar on the right. The KLIP parameters used for each reduction are given in Table~\ref{tbl:gpi-log}.    \label{fig:bka}}
\end{figure}

\begin{deluxetable*}{ccccccccc}
\tablewidth{0pt}
\tablecaption{Relative astrometry between HR~2562 A and HR~2562 B\label{tbl:astrometry}}
\tablehead{ \colhead{UT Date} & \colhead{MJD} & \colhead{Instrument} & \colhead{Filter} & \colhead{Plate scale} & \colhead{North offset} & \colhead{$\rho$} & \colhead{$\theta$} & \colhead{Ref.} \\ 
& & & & (mas\,px$^{-1}$) & (deg) & (mas) & (deg) & }
\startdata
2016 Jan 25 & 57412.1335 & Gemini-S/GPI & $H$ & $14.161\pm0.021$ & $0.21\pm0.23$ & $615.05\pm1.76$ & $297.83\pm0.27$ & 1\\
2016 Jan 28 & 57415.1731 & Gemini-S/GPI & $K_1$ & $14.161\pm0.021$ & $0.21\pm0.23$ & $612.56\pm1.46$ & $297.76\pm0.25$ & 1\\
2016 Jan 28 & 57415.2002 & Gemini-S/GPI & $K_2$ & $14.161\pm0.021$ & $0.21\pm0.23$ & $613.47\pm1.28$ & $298.09\pm0.24$ & 1\\
2016 Feb 25 & 57443.0343 & Gemini-S/GPI & $K_2$ & $14.161\pm0.021$ & $0.21\pm0.23$ & $616.09\pm1.23$ & $297.71\pm0.24$ & 1\\
2016 Feb 28 & 57446.0951 & Gemini-S/GPI & $J$ & $14.161\pm0.021$ & $0.21\pm0.23$ & $616.44\pm1.37$ & $297.78\pm0.25$ & 1\\
2016 Dec 12 & 57734.2647 & VLT/SPH-IRD & $H$ & $12.251\pm0.009$ & $-1.808\pm0.043$ & $637.80\pm6.40$ & $297.81\pm0.54$ & 2\\
2016 Dec 17 & 57739.3123 & Gemini-S/GPI & $K_1$ & $14.161\pm0.021$ & $0.32\pm0.15$ & $639.89\pm1.27$ & $297.98\pm0.17$ & 1\\
2017 Feb 07 & 57791.1114 & VLT/SPH-IRD & $H$ & $12.251\pm0.009$ & $-1.712\pm0.058$ & $644.00\pm2.30$ & $297.82\pm0.19$ & 2\\
2017 Feb 13 & 57797.0517 & Gemini-S/GPI & $H$ & $14.161\pm0.021$ & $0.32\pm0.15$ & $644.20\pm1.22$ & $298.23\pm0.16$ & 1\\
2017 Sep 29 & 58025.3826 & VLT/SPH-IRD & $K_1$ & $12.267\pm0.009$ & $-1.735\pm0.043$ & $661.20\pm1.30$ & $297.97\pm0.16$ & 2\\
2017 Sep 29 & 58025.3826 & VLT/SPH-IRD & $K_2$ & $12.263\pm0.009$ & $-1.735\pm0.043$ & $658.90\pm1.60$ & $298.08\pm0.17$ & 2\\
2017 Nov 29 & 58086.3110 & Gemini-S/GPI & $H$ & $14.161\pm0.021$ & $0.28\pm0.19$ & $664.82\pm2.04$ & $298.37\pm0.24$ & 1\\
2018 Jan 31 & 58149.1962 & Gemini-S/GPI & $K_2$ & $14.161\pm0.021$ & $0.28\pm0.19$ & $669.44\pm1.24$ & $298.55\pm0.20$ & 1\\
2018 Mar 10 & 58187.0475 & Gemini-S/GPI & $Y$ & $14.161\pm0.021$ & $0.28\pm0.19$ & $670.84\pm2.83$ & $298.74\pm0.26$ & 1\\
2018 Nov 19 & 58441.3261 & Gemini-S/GPI & $H$ & $14.161\pm0.021$ & $0.45\pm0.11$ & $685.76\pm1.25$ & $298.89\pm0.13$ & 1\\
\enddata
\tablerefs{(1) - this work; (2) - \citet{Maire:2018ch}.}
\end{deluxetable*}

Although GPI can routinely achieve high Strehl ratios on bright stars, it does not offer perfect suppression of starlight. Within about $1\arcsec$ the images are dominated by residual quasi-static speckles caused by AO fitting errors and non-common path aberrations. This residual light often has a pronounced azimuthal asymmetry in the direction of the jet stream \citep{Madurowicz:2018kw}, which can result in a significant azimuthal dependence on the achieved contrast. We took advantage of both ADI and spectral differential imaging \citep[SDI,][]{Smith:1987ea,Racine:1999gq}) to subtract these quasi-static speckles from each slice within each data cube. We used \texttt{pyKLIP} \citep{Wang:2015th}, a Python implementation of the Karhunen--Lo\`{e}ve image projection algorithm \citep[KLIP,][]{Soummer:2012ig,Pueyo:2015cx}), to model and subtract the residual starlight within each image. Due to the distorting effects of this algorithm on the PSF of the companion within the PSF-subtracted image, we used the forward-model based Bayesian KLIP-FM astrometry package \citep[BKA,][]{Wang:2016gl}) to model the effect of KLIP on the PSF of the companion (see Figure \ref{fig:bka}). We used the average of the four satellite spots as the instrumental PSF, and low-throughput channels were excluded.

We used the same approach for the BKA forward-modelling as in \cite{DeRosa:2020gy}. The PSF subtraction and forward model was calculated within a single annulus centered on the star with a width of either 16\,px ($Y$, $J$, $H$) or 20\,px ($K_1$, $K_2$). The radius of the annulus was selected to center the companion between the inner and outer edge. We explored the effect of varying two of the main tunable parameters: the movement criteria, $m$, that defines the minimum number of pixels an astrophysical source needs to have moved by before an image can be included within the reference PSF library, and the number of KL modes, $n_{\rm KL}$, used to reconstruct the stellar PSF. We measured astrometry using BKA for each combination of these parameters. The forward model of the PSF was compared to the companion within a small $11\times11$\,px box ($15\times15$\,px for $K_1$ and $K_2$). Posterior distributions of the position and flux of the companion, and the correlation length scale (a nuisance parameter to marginalize over) were sampled using the Markov chain Monte Carlo (MCMC) affine-invariant sampler within the \texttt{emcee} package \citep{ForemanMackey:2013io}. We advanced 100 walkers that were initialized near the expected parameter values for 1000 steps, discarding the first 200 as burn-in. 

Using this set of measurements we investigated the effect of the KLIP parameters on the companion astrometry. Small values of $m$ and large values of $n_{\rm KL}$ can cause significant self-subtraction or over-subtraction of the companion, an effect most pronounced at the longer wavelengths where the PSF is larger. We found a significant correlation between these two parameters and the position of the companion for the two $K_2$ datasets, most pronounced in the 2018 Jan 31 epoch. For the other epochs at shorter wavelengths no significant correlation was observed. The adopted KLIP parameters are listed in Table~\ref{tbl:astrometry}. A low value of $n_{\rm KL}$ was adopted given the relative brightness of the companion, and a high value of $m$ to minimize self-subtraction especially at longer wavelengths. The posterior distributions of the companion position measured using the adopted parameter set were combined with the instrument plate scale and orientation calibrations from \cite{DeRosa2020:kd} to yield on-sky relative astrometric measurements for each epoch (Table~\ref{tbl:astrometry}). An image of the instrumental PSF, forward model, companion, and residuals are shown for each dataset in Figure~\ref{fig:bka}.

\section{Result} 
\label{section:results}

\subsection{ALMA image}
\label{sec:ALMA image}
As shown in Figure \ref{fig:Data}, the disk is clearly detected and well resolved along the major axis at 870\,$\mu$m. 
Placing a beam-sized aperture around the brightest pixel in the image, we evaluate a peak signal to noise of $\sim$7 in the ALMA image.
Owing to spatial filtering by the interferometer,
the reconstructed image has negative pixels surrounding the disk. This makes it difficult to measure the total flux of the disk directly from the ALMA image. To correct for this, we instead report a total flux of 3.3mJy from integrating the best fit model image (see section 4.1). 
These observations extend the system's SED to a longer wavelength than all prior observations, most notably beyond the range of {\sl Herschel}. 
With an angular resolution that is $\approx$6 times higher than {\sl Herschel}'s, the disk geometry is much more clearly apparent than in previous studies \citep{Moor:2015kf}. The highly elongated structure reveals the disk to be nearly edge-on. Fitting a 2D Gaussian to the map results in a FWHM of 8\farcs17 along the major axis and 0\farcs97 along the minor axis of the disk. The aspect ratio of the two corresponds to an inclination of $\approx$\,83\degr, assuming a thin disk. We notice that the FWHM of the minor axis is marginally smaller than the size of the beam, indicating the disk is unresolved along the minor axis. Therefore, the $\approx$83\degr\ inclination we inferred from the aspect ratio is a strict lower limit. 

To verify the alignment between the detected emission and the location of the star, we use the absolute astrometry information of the ALMA map, which is known to a precision of about 0\farcs1 based on the pointing calibration. 
The position of the star is retrieved from the Gaia EDR3 Release, taking into account its proper motion and the epoch of the ALMA observation \citep{GaiaCollaboration:2021ev}. The star and the center of the disk, as defined by our 2D Gaussian fit,  are within 0\farcs07 of each other, i.e., consistent with a symmetric disk about the star (see Figure \ref{fig:Data}).
The location of the star and the center of the disk are shown in right panel of Figure \ref{fig:Data}.

\begin{figure}
   \centering
   \includegraphics[width=0.5\textwidth]{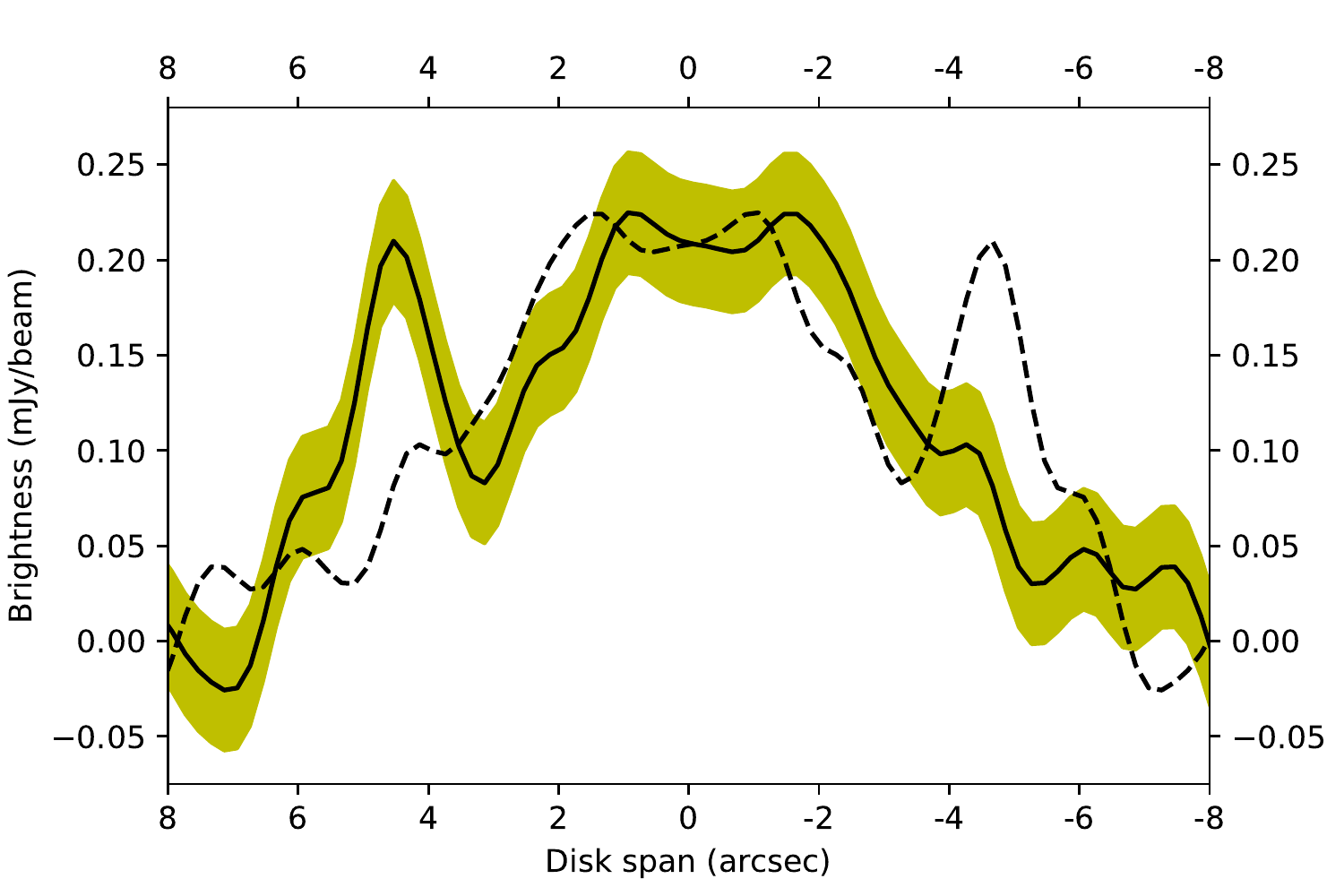} 
   \caption{870\,$\mu$m brightness profile along the disk major axis. The yellow shade indicates the 1\,$\sigma$ uncertainty as determined from the surrounding background areas. The dashed line represent a horizontally reflected brightness profile about the location of the star.  
   \label{fig:bright_prof}}
\end{figure}

From the image, a clump-like structure can be observed in the southeast side of the disk. This clump also manifests in the surface brightness profile of the disk shown in Figure {\ref{fig:bright_prof}}, where we see a peak between 2'' and 6''.  We show a horizontally flipped profile in dashed line in  Figure {\ref{fig:bright_prof}} , from which we can see that the peak stands at a marginally significant ($\sim 2.5 \sigma$) difference measuring from the height of the brightness profile at the horizontally reflected location of the peak. There are a few plausible explanations for this feature. 1) It is due to the random noise of the image; 2) The trough is a real feature that represents a  real physical underdensity, e.g. a gap or spiral in the disk, thus making the disk asymmetrical;  3) The trough was the result of a contamination from an unrelated point source, presumably a background submillimeter galaxy.  Whilst a physical disk origin cannot be ruled out, the signal-to-noise ratio means cannot say with confidence that the feature is real.  In addition, although it is possible that there might be previously unseen background galaxy, the chance of it aligning so well with the disk is low. 
We will therefore model the disk with a symmetric profile in the analysis that follows, but note that future observations may better determine the nature of the trough.

\subsection{Orbit of the brown dwarf companion}
\label{sec:orbit result}
\subsubsection{Unconstrained fit}
\label{sec:orbit-visual}

The orbit of the companion was fitted to the measured astrometry in Table~\ref{tbl:astrometry} using a modified version of the procedure described in \cite{DeRosa:2020gy}. The visual orbit can be described using the standard Campbell elements; the period ($P$), semi-major axis ($a$), eccentricity ($e$), inclination ($i$), argument of periastron ($\omega$), position angle of the ascending node ($\Omega$), and the time of periastron ($T_0$). Here, we use the standard convention that $\Omega$ refers to the position angle that the companion passes through the plane tangent to the sky at the location of HR 2562, moving away from the observer. We substituted two of these elements when fitting the orbit of HR~2562. $P$ was replaced with the total system mass ($M_{\rm total}$) as the period is currently unconstrained given the measurements, and $T_0$ was replaced with the dimensionless parameter $\tau$ describing the time of the next periastron passage in fractions of the orbital period since MJD 57412.1335, the start of the astrometric record. In addition to these seven elements we also required the distance to the star, represented by the parallax ($\varpi$), in order to link the angular semi-major axis to the system mass to derive the orbital period.

We used the parallel-tempered affine-invariant Markov chain Monte Carlo (MCMC) sampler within the {\tt emcee} package \citep{ForemanMackey:2013io} to sample the posterior distribution of these eight parameters describing the visual orbit of HR~2562 B. We performed the sampling using the parameters $\log a$ and $\cos i$ to enforce a log-uniform prior probability density function (PDF) on $a$ and a sine prior PDF on $i$ rather than computing the prior probability at each step. Gaussian priors were used for $M_{\rm total}$ \citep[$1.31\pm0.13$\,$M_\odot$;][]{Moor:2015kf}, under the assumption that $M_2\ll M_1$, and $\varpi$ \citep[$29.4738\pm0.0185$\,mas;][]{GaiaCollaboration:2021ev}. We advanced 512 chains at each of 16 different temperatures for one million steps. In the parallel-tempered framework the lowest temperature chains sample the posterior probability distribution, while the highest temperature chains sample the prior probability distribution. We saved every hundredth sample of each chain to disk, and conservatively discarded the first half of each as a ``burn-in''.

\begin{figure*}[]
   \centering
   \includegraphics[width=\textwidth]{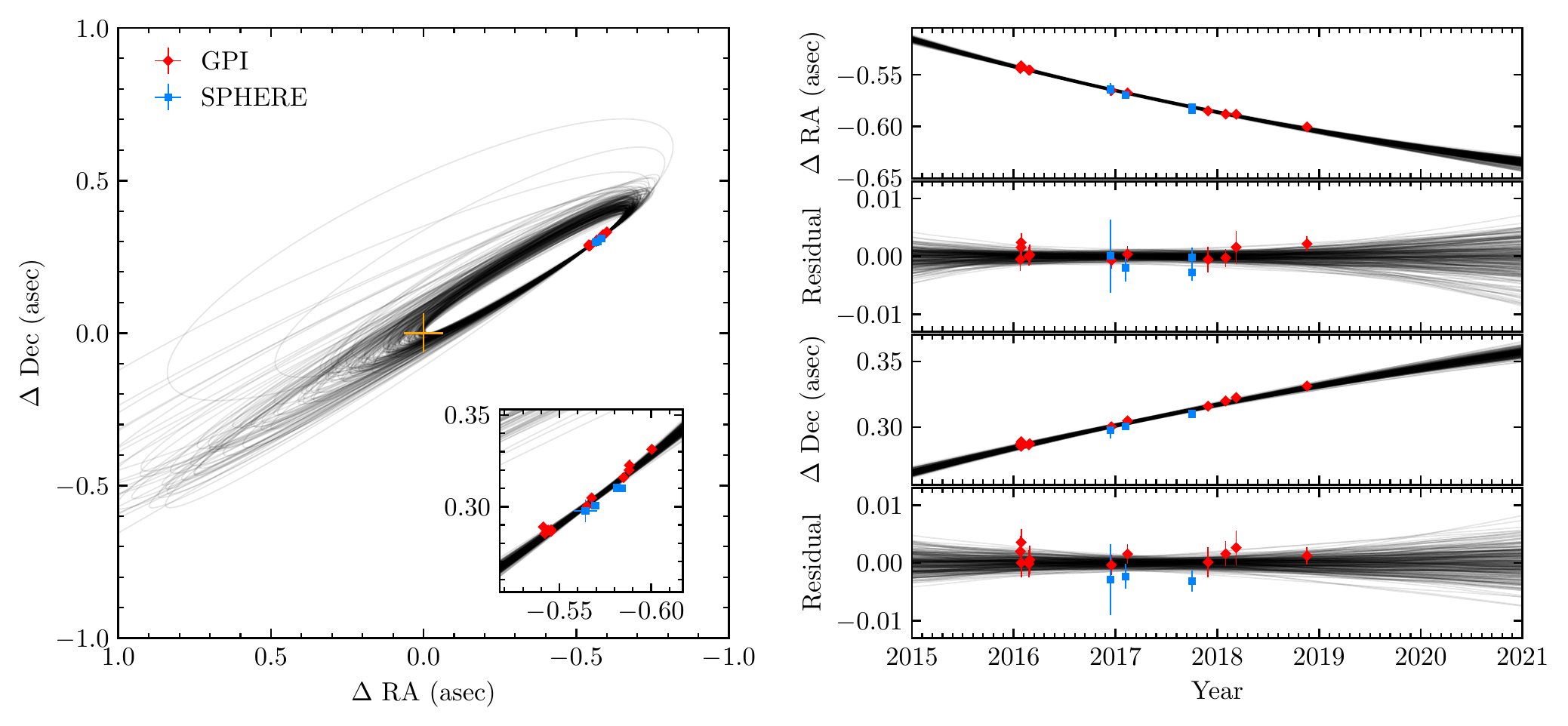} 
   \caption{(left panel) Relative astrometry of HR~2562 B relative to HR~2562 from GPI (red) and SPHERE (blue). The location of the host star is indicated by the orange cross. Two hundred orbits drawn from the MCMC analysis (black curves) are also shown. (right panel) Offset between HR 2562 and the companion, and the residual compared with the median orbit, in the right ascension (top) and declination (bottom) directions.}
   \label{fig:orbit}
\end{figure*}

The median and one-sigma confidence intervals for the fitted and derived parameters are shown in Table~\ref{tbl:orbit}, along with the set of parameters describing the maximum likelihood (minimum $\chi^2$) and maximum probability orbit. A random selection of visual orbits consistent with the relative astrometry are shown in Figure \ref{fig:orbit}. The posterior PDFs for a subset of the fitted and derived parameters are shown in Figure \ref{fig:orbit-corner1} (black contours). We found a strong anti-correlation between the eccentricity and the inclination of the orbit; less eccentric orbits tend to have a more edge-on configuration ($i\sim84$\,\degr). Although very high eccentricities are seemingly preferred based on the shape of the PDF, the eccentricity of the orbit is not yet well constrained and still dependent on the input priors. We discuss this further in section \ref{low eccentricity}.

\subsubsection{Co-planar scenario}
\label{subsec:coplanar}
The results regarding the orbit of the companion presented so far make no assumptions regarding the alignment of the orbital plane with that of the debris disk resolved in the ALMA observations presented in Section~\ref{sec:alma-obs}. We repeated the orbit fitting procedure described above with an additional prior on the inclination ($i$) and position angle of the ascending node ($\Omega$) to investigate the properties of orbits consistent with both the measured astrometry and a near co-planar configuration with respect to the external debris disk. We used Gaussian kernel density estimation (KDE) to construct a prior probability density function on the orbital inclination ($i$) and position angle ($\Omega$) from the disk fitting MCMC samples (Sec.~\ref{sec:disk_modeling}). We used the \texttt{scipy.gaussian\_kde} function with Scott's method for the estimator bandwidth calculation. The resulting two-dimensional prior distribution is shown in Figure \ref{fig:orbit-kde}, along with the two marginalized distributions. The Gaussian KDE is a good match to the posterior distributions describing the disk geometry from the ALMA observations presented in Section~\ref{sec:disk_modeling}. The additional term in the prior probability was calculated using the Gaussian KDE and the value of $i$ and $\Omega$ at each step in the MCMC process. We accounted for the ambiguity in these two parameters when fitting to the ALMA data by evaluating the Gaussian KDE at four possible combinations of $i$ and $\Omega$: ($i$, $\Omega$), ($i$, $\Omega+\pi$), ($\pi - i$, $\Omega$), and ($\pi - i$, $\Omega + \pi$). The maximum of these four values was used as the prior probability at this step for this combination of $i$ and $\Omega$.

The median and 1-sigma credible intervals for the posterior distributions on the fitted and derived parameters from this analysis are given in Table~\ref{tbl:orbit}, in addition to the maximum likelihood and maximum probability orbit. The posterior distributions of a subset of the fitted and derived parameters are compared to those calculated without the additional prior in Figure \ref{fig:orbit-corner1}. The full corner plot is presented in Appendix \ref{append:Full orbit fit}. The additional prior restricts the range of orbital eccentricities that are consistent with the measured astrometry in that very eccentric orbits ($e\gtrsim0.8$) require increasingly misaligned configurations with respect to the outer disk when considering the general fit. The other orbital parameters are not significantly affected by the prior, in particular the apoastron distance is unchanged due to the anti-correlation between the orbital eccentricity and semi-major axis.

\subsubsection{Dynamical mass constraint}
\label{sec:orbit-hg}
We used the relative astrometry between the star and companion presented here and absolute astrometric measurements of the photocenter from the {\it Hipparcos} and {\it Gaia} satellites to constrain the mass of the companion. With a sufficiently high mass for the companion, the photocenter of the system will be perturbed from linear motion through space due to the orbit of the star around the barycenter of the system. For {\it Hipparcos} we used the intermediate astrometric data (IAD), i.e., the residuals along the scan direction from the best fit astrometric solution \citep{vanLeeuwen:2007du}. These astrometric measurements of HR~2562 were best fit with a non-standard ``stochastic'' solution where the uncertainties on the individual measurements are inflated to reduce the goodness of fit statistic below an acceptable threshold for the final catalogue \citep{vanLeeuwen:2007dc}. The need for this type of solution can be attributed to unresolved orbital motion on a time-scale much shorter than the three year duration of the mission, or to remaining modelling noise from the construction of the catalogue \citep{vanLeeuwen:2007du}.

For {\it Gaia} we used the astrometric parameters and correlation coefficients reported in the Early Data Release 3 catalogue (EDR3; \citealp{GaiaCollaboration:2021ev}). We compared the various goodness of fit metrics reported in the EDR3 catalogue for HR~2562 to those of stars with a similar brightness. As illustrated in Appendix\,\ref{append:gaia}, HR~2562 appears relatively well behaved compared to these other sources. We also collected the timings of the individual {\it Gaia} measurements using the {\it Gaia} Observation Forecast Tool\footnote{\url{https://gaia.esac.esa.int/gost/}}, excluding those taken during satellite down-times reported in \citet{Lindegren:2021jk}.

The procedure used in \citet{DeRosa:2020jc} for a joint fit of absolute astrometry and radial velocity measurements was adapted to fit the absolute and relative astrometric measurement available for HR~2562, as well as updating it to account for the change in reference epoch between {\it Gaia} DR2 and EDR3. The model that simultaneously describes the orbit of the companion around HR~2562 and the orbit of the photocenter around the barycenter of the system consists of 14 parameters. Seven parameters are  used in the orbit fit described in Section~\ref{sec:orbit-visual} ($\log a$, $\cos i$, $e$, $\omega$, $\Omega$, $\tau$, $\varpi$), two describe the masses of the components ($M_1$, $\log M_2$), four describe offsets between the {\it Hipparcos} catalogue astrometry and the barycenter position and proper motion at 1991.25 ($\Delta \alpha^\star$, $\Delta \delta$, $\Delta \mu_{\alpha^\star}$, $\Delta \mu_\delta$), and one is an error inflation term applied in quadrature to the {\it Hipparcos} IAD ($\epsilon_\Lambda$). To calculate the semi-major axis of the photocenter orbit we used an empirical mass-magnitude relationship to determine the relative fluxes of the two components in the {\it Hipparcos} and {\it Gaia} passbands when $M_2 > 0.077$\,$M_{\odot}$ \citep{Pecaut:2013ej}. For lower masses we assumed that the companion is emitting no flux, and that the photocenter is coincident with the location of the host star.

We used the same MCMC sampler as in Section~\ref{sec:orbit-visual} to sample the posterior distributions of the fourteen parameters in this model. We used Gaussian priors for the parallax ($\varpi=29.4738\pm0.0185$\,mas) and for the mass of the primary ($M_1=1.31\pm0.13$\,$M_{\odot}$), uniform priors were used for the remaining parameters. We advanced 512 chains at each of 16 different temperatures for one million steps, saving every tenth sample to disk. The first half of each chain was discarded as a ``burn-in''. 

\begin{figure}[]
   \centering
   \includegraphics[width=0.5\textwidth]{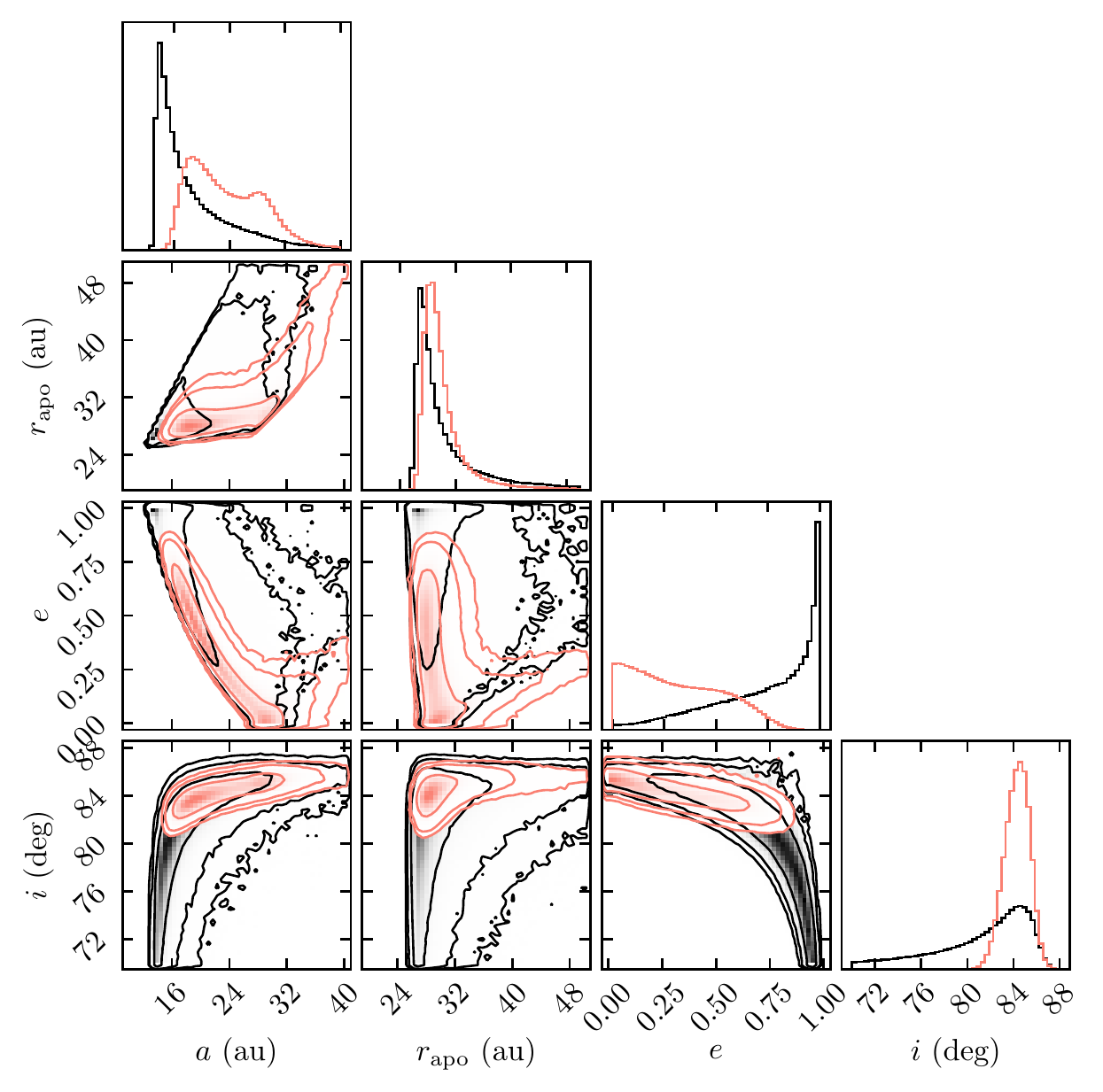} 
   \caption{Posterior distributions and associated covariances of the semi-major axis ($a$), periastron distance ($r_{\rm apo}$), eccentricity ($e$) and inclination ($i$) from an orbit fit with (red) and without (black) a prior probability on the inclination and position angle ($\Omega$) derived from the ALMA disk fitting.  Contours represent the 1, 2, and 3\,$\sigma$ confidence levels. Limiting the orbit to a co-planar configuration removes the tail of extremely high eccentricity orbits. \label{fig:orbit-corner1}}
\end{figure}

\begin{figure}[]
   \centering
   \includegraphics[width=0.5\textwidth]{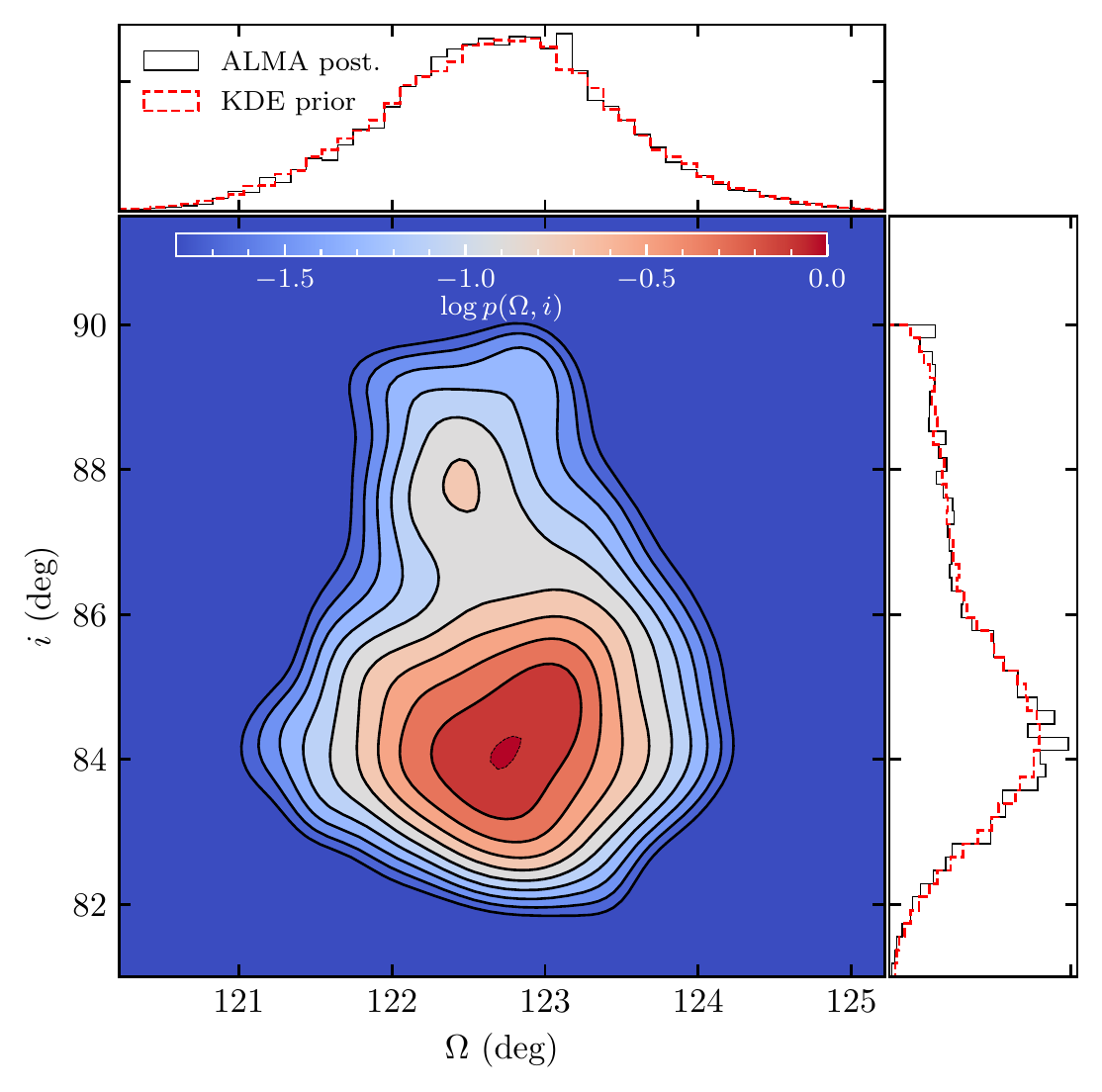} 
   \caption{Normalized joint prior probability on the orbital inclination $i$ and position angle $\Omega$ estimated using a Gaussian KDE for the co-planar scenario (main panel). Solid contours are spaced in 0.2\,dex increments from $-1.8$ to $-0.2$, the inner dashed contour is at $-0.01$. Marginalized distributions are shown in the top and right panels for the posterior distribution estimated from the ALMA data (black solid histogram, Section~\ref{sec:disk_modeling}) and for the prior distribution used in the orbit fit (red dashed histogram).}
   \label{fig:orbit-kde}
\end{figure}

\begin{figure}[]
   \centering
   \includegraphics[width=0.5\textwidth]{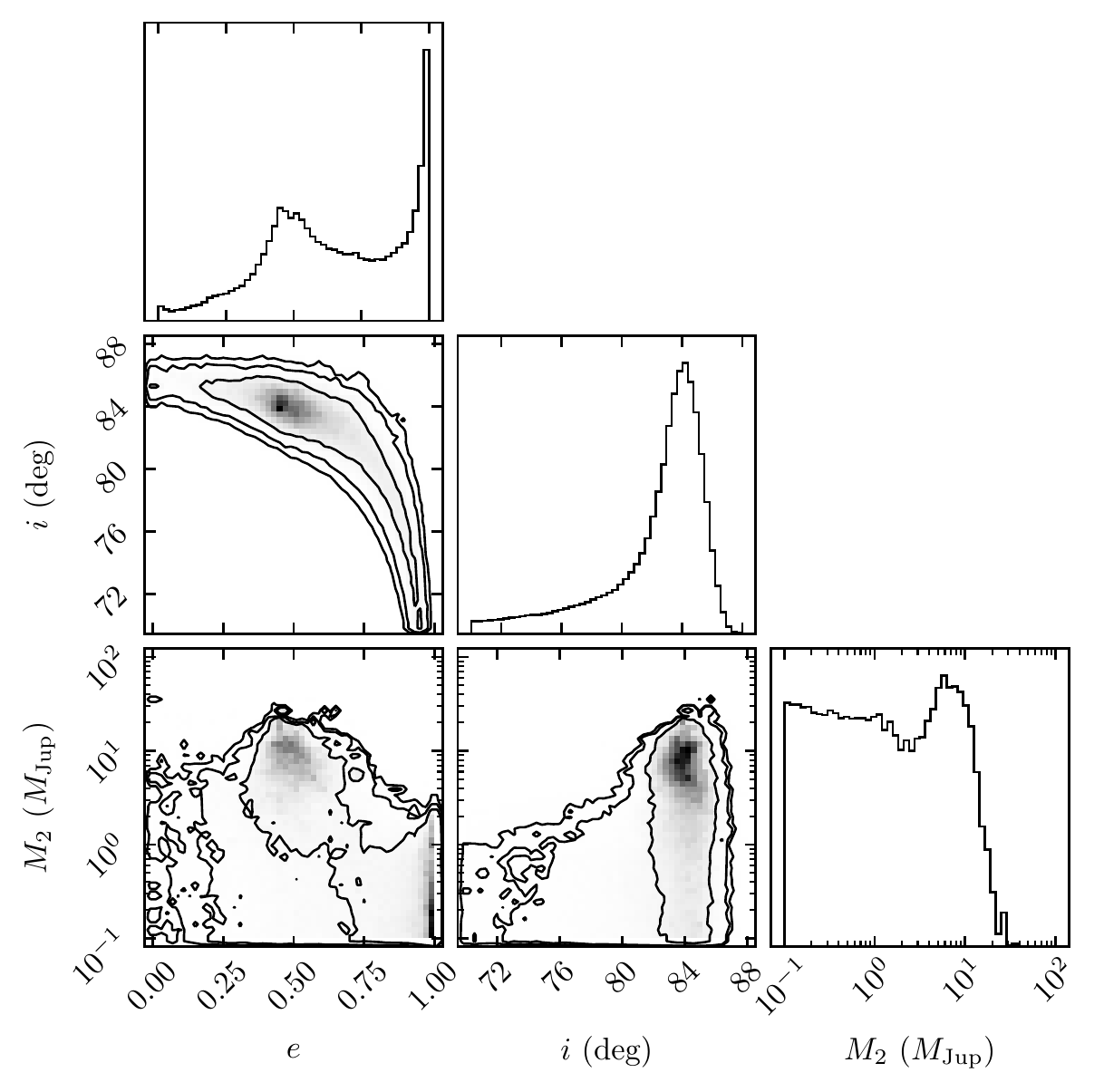} 
   \caption{Posterior distributions and associated covariance of the orbital eccentricity ($e$), inclination ($i$), and companion mass ($M_2$) from a joint fit to the relative and absolute astrometry.  Contours represent the 1, 2, and 3\,$\sigma$ confidence intervals.}
   \label{fig:orbit-corner2}
\end{figure}

\begin{figure}[]
   \centering
   \includegraphics[width=0.5\textwidth]{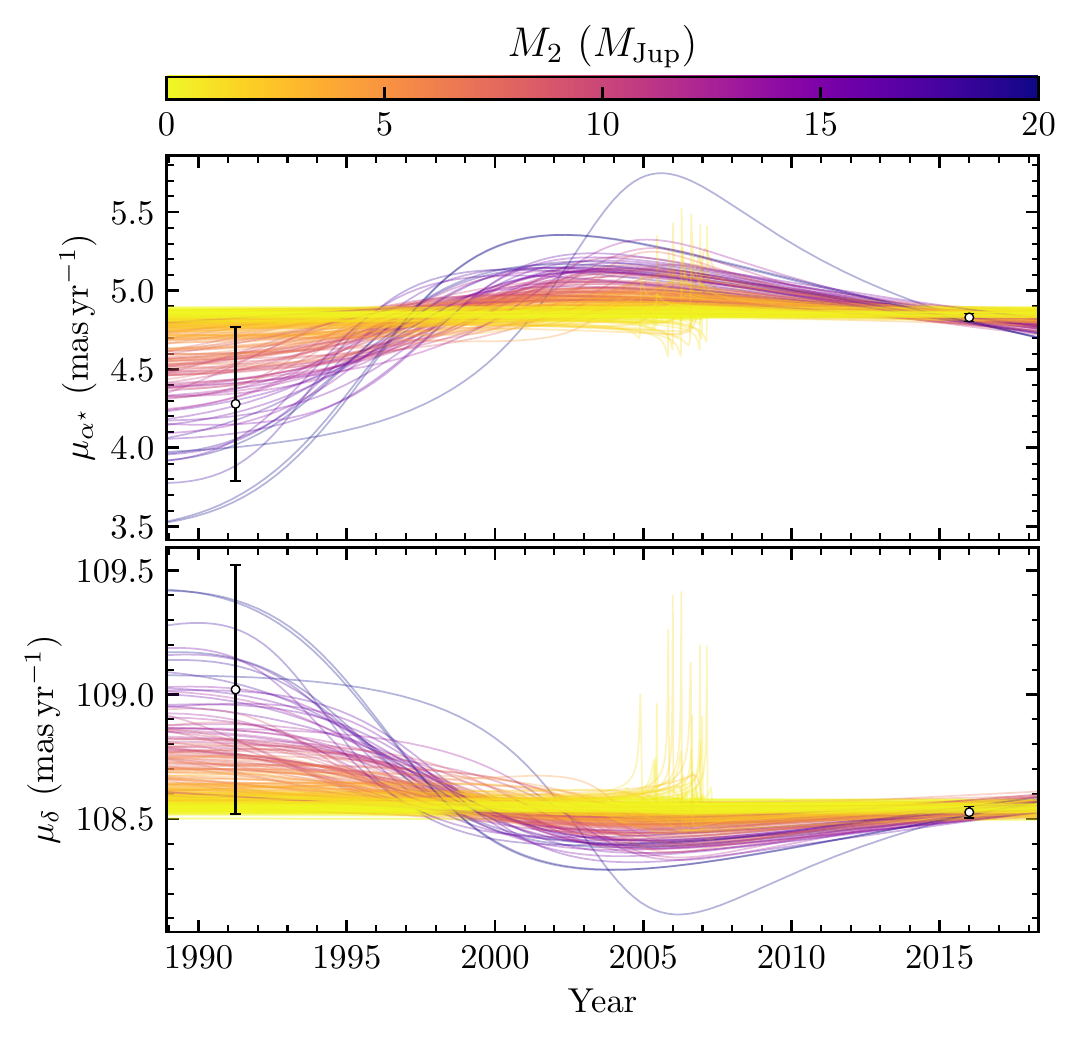} 
   \caption{Apparent proper motion of the photocenter of the HR 2562 system in the right ascension (top) and declination (bottom) directions for two hundred orbits drawn from the MCMC analysis presented in Section~\ref{sec:orbit-hg} that are consistent with the absolute and relative astrometry. The tracks are colored by the mass of the companion for each orbit. The astrometric measurements from the {\it Hipparcos} and {\it Gaia} catalogues are plotted.}
   \label{fig:orbit-pmfit}
\end{figure}

The median and one-sigma confidence intervals for the fitted and derived parameters are shown in Table~\ref{tbl:orbit}, along with the set of parameters describing the maximum likelihood and maximum probability orbit. The covariance between the eccentricity, inclination, and mass of the companion is shown in Figure \ref{fig:orbit-corner2}. The eccentricity and inclination exhibit a similar anti-correlation that is seen for the visual orbit fit (Figure \ref{fig:orbit-corner1}), albeit with a slight enhancement of orbits with a moderate eccentricity ($e\sim0.5$). The shape of the $M_2$ posterior probability distribution (Figure \ref{fig:orbit-corner2}, lower right panel) clearly demonstrates that we do not yet have a statistically significant measurement of the mass of the companion. Instead, we can only place an upper limit of $M_2<18.5$\,$M_{\rm Jup}$ at the 99\% confidence level given our assumptions regarding the prior probability distribution. It should be noted that the exact number, while prior-dependent, is expected to be rather precise given the sharp truncation of the posterior. Below this value the posterior probability distribution is not significantly different from that of the prior probability distribution that was uniform in $\log M_2$. The proper motion of the HR~2562 photocenter between the {\it Hipparcos} and {\it Gaia} missions is shown in Figure \ref{fig:orbit-pmfit}. The proper motion measurements are consistent with companion masses $\lesssim20$\,$M_{\rm Jup}$; higher-mass companions would have caused a larger amplitude astrometric reflex motion between the two epochs, inconsistent with the observations.

\subsubsection{The case for low eccentricity solution}
\label{low eccentricity}
As shown here, the limited coverage of the companion's orbit still leaves a broad range of possible orbits. In short, the data are consistent with both low- to moderate-eccentricity orbits that are roughly coplanar with the debris disk and high-eccentricity orbits at a significantly different inclination. It is worth noting that, with such limited orbital phase coverage, Keplerian fits are notoriously subject to eccentricity and inclination biases \citep[e.g.,][]{Lucy2014, FerrerChavez2021} and they remain highly sensitive to the MCMC priors \citep{Pearce2015, ONeill2019}. These effects are highly sensitive to the exact orbital phase coverage and system viewing geometry, so they are hard to evaluate and correct for. While only more extensive astrometric coverage will solve both issues, it is worth taking an hollistic approach of the results presented here. In particular, a high-eccentric and misaligned companion would most likely result in an eccentric disk, through apsidal alignment, and/or a significant warp in the system. The ALMA map presented here does not provide strong evidence for either phenomenon, suggesting that a lower eccentricity and near coplanarity should be preferred. 

As an additional test, we performed an orbital fit where we forced $e=0$. The resulting fit is slightly poorer, with $\chi^2 = 21.9$ (to be compared with $\chi^2 = 20.0$ for the eccentric fit). Given 15 two-dimensional datapoints and 6 (8) free parameters for the circular (eccentric) fit, both the Bayesian Information Criterion and the Aikake Information Criterion indicate that the data are sufficiently well fit by the circular fit. Since it is unlikely that the orbital is exactly circular, we will use the result of the full Keplerian fit in the remainder of this study, but we consider that both physical arguments and the circular fit provide suggestive evidence against the high-eccentricity orbital solutions.

\begin{deluxetable*}{ccccccccccc}
\tablewidth{0pt}
\tablecaption{Orbital elements for HR 2562 B\label{tbl:orbit}}
\tablehead{
\colhead{Parameter} & \colhead{Unit} & \multicolumn{3}{c}{Visual} & \multicolumn{3}{c}{Visual + Prior} & \multicolumn{3}{c}{Visual + Absolute}\\
& & \colhead{Range} & \colhead{max. $\mathcal{L}$} & \colhead{max. $\mathcal{P}$} & \colhead{Range} & \colhead{max. $\mathcal{L}$} & \colhead{max. $\mathcal{P}$} & \colhead{Range} & \colhead{max. $\mathcal{L}$} & \colhead{max. $\mathcal{P}$}}
\startdata
\multicolumn{11}{c}{Fitted parameters}\\
\hline
$\log a$ & [arc sec] & $-0.305_{-0.078}^{+0.180}$ & $-0.286$ & $-0.401$ & $-0.170_{-0.099}^{+0.112}$ & $-0.343$ & $-0.260$ & $-0.252_{-0.116}^{+0.115}$ & $-0.234$ & $-0.204$ \\
$\cos i$ & \nodata & $0.165_{-0.073}^{+0.306}$ & $0.110$ & $0.531$ & $0.098_{-0.017}^{+0.020}$ & $0.131$ & $0.114$ & $0.125_{-0.036}^{+0.211}$ & $0.089$ & $0.103$ \\
$e$ & \nodata & $0.79_{-0.35}^{+0.18}$ & $0.58$ & $0.98$ & $0.29_{-0.21}^{+0.28}$ & $0.67$ & $0.52$ & $0.63_{-0.23}^{+0.32}$ & $0.45$ & $0.55$ \\
$\omega$ & deg & $202.2_{-26.6}^{+42.7}$ & $210.4$ & $186.2$ & $179.8_{-17.3}^{+69.1}$ & $182.5$ & $170.9$ & $207.2_{-30.1}^{+34.6}$ & $216.5$ & $230.9$ \\
$\Omega$ & deg & $299.2_{-3.8}^{+3.1}$ & $298.7$ & $298.0$ & $302.1_{-0.8}^{+0.8}$ & $300.5$ & $302.3$ & $299.5_{-2.7}^{+2.5}$ & $299.3$ & $298.5$ \\
$\tau$ & \nodata & $0.794_{-0.040}^{+0.079}$ & $0.780$ & $0.785$ & $0.696_{-0.089}^{+0.078}$ & $0.716$ & $0.685$ & $0.795_{-0.057}^{+0.075}$ & $0.785$ & $0.841$ \\
$\varpi$ & mas & $29.474_{-0.018}^{+0.018}$ & $29.464$ & $29.476$ & $29.474_{-0.018}^{+0.018}$ & $29.463$ & $29.473$ & $29.474_{-0.018}^{+0.018}$ & $29.507$ & $29.474$ \\
$M_{\rm Total}$ & $M_{\odot}$ & $1.34_{-0.12}^{+0.13}$ & $1.67$ & $1.33$ & $1.34_{-0.12}^{+0.12}$ & $1.67$ & $1.37$ & \nodata & \nodata & \nodata \\
$M_{1}$ & $M_{\odot}$ & \nodata & \nodata & \nodata & \nodata & \nodata & \nodata & $1.34_{-0.12}^{+0.13}$ & $1.44$ & $1.36$ \\
$\log M_{2}$ & [$M_{\rm Jup}]$ & \nodata & \nodata & \nodata & \nodata & \nodata & \nodata & $-0.40_{-1.07}^{+1.14}$ & $1.07$ & $1.01$ \\
$\Delta \alpha^\star$ & mas & \nodata & \nodata & \nodata & \nodata & \nodata & \nodata & $0.59_{-0.43}^{+0.66}$ & $1.90$ & $1.75$ \\
$\Delta \delta$ & mas & \nodata & \nodata & \nodata & \nodata & \nodata & \nodata & $0.06_{-0.51}^{+0.39}$ & $-0.72$ & $-0.48$ \\
$\Delta \mu_{\alpha^\star}$ & mas\,yr$^{-1}$ & \nodata & \nodata & \nodata & \nodata & \nodata & \nodata & $0.56_{-0.11}^{+0.02}$ & $0.35$ & $0.37$ \\
$\Delta \mu_\delta$ & mas\,yr$^{-1}$ & \nodata & \nodata & \nodata & \nodata & \nodata & \nodata & $-0.46_{-0.02}^{+0.06}$ & $-0.35$ & $-0.36$ \\
$\sigma_\Lambda$ & mas & \nodata & \nodata & \nodata & \nodata & \nodata & \nodata & $2.35_{-0.24}^{+0.26}$ & $2.23$ & $2.25$ \\
\hline
\multicolumn{11}{c}{Derived parameters}\\
\hline
$P$ & yr & $60.1_{-14.7}^{+51.2}$ & $56.8$ & $42.9$ & $95.4_{-28.6}^{+45.1}$ & $46.8$ & $68.8$ & $71.5_{-23.2}^{+35.7}$ & $72.9$ & $83.5$ \\
$a$ & arc sec & $0.496_{-0.082}^{+0.255}$ & $0.517$ & $0.397$ & $0.676_{-0.138}^{+0.198}$ & $0.454$ & $0.549$ & $0.560_{-0.131}^{+0.169}$ & $0.583$ & $0.625$ \\
$a$ & au & $16.8_{-2.8}^{+8.7}$ & $17.6$ & $13.5$ & $22.9_{-4.7}^{+6.7}$ & $15.4$ & $18.6$ & $19.0_{-4.4}^{+5.7}$ & $19.8$ & $21.2$ \\
$r_{\rm apo}$ & au & $28.8_{-1.9}^{+8.4}$ & $27.7$ & $26.6$ & $29.3_{-1.5}^{+3.3}$ & $25.7$ & $28.4$ & $29.2_{-2.1}^{+7.2}$ & $28.6$ & $32.9$ \\
$i$ & deg & $80.5_{-18.6}^{+4.2}$ & $83.7$ & $57.9$ & $84.4_{-1.2}^{+1.0}$ & $82.5$ & $83.5$ & $82.8_{-12.5}^{+2.0}$ & $84.9$ & $84.1$ \\
$T_0$ & yr & $2005.0_{-4.9}^{+1.6}$ & $2003.5$ & $2006.8$ & $1991.0_{-20.2}^{+7.6}$ & $2002.8$ & $1994.4$ & $2003.3_{-6.1}^{+2.9}$ & $2000.4$ & $2002.8$ \\
$M_{2}$ & $M_{\rm Jup}$ & \nodata & \nodata & \nodata & \nodata & \nodata & \nodata & $<18.5$ & $11.67$ & $10.28$ \\
\hline
\multicolumn{11}{c}{Goodness of fit}\\
\hline
$\chi^2_{\rho}$ & \nodata & $6.79_{-1.66}^{+2.96}$ & $3.54$ & $4.34$ & $10.61_{-2.49}^{+3.19}$ & $3.92$ & $7.91$ & $6.88_{-1.71}^{+3.00}$ & $4.23$ & $4.21$ \\
$\chi^2_{\theta}$ & \nodata & $17.84_{-1.03}^{+2.28}$ & $16.48$ & $16.46$ & $17.80_{-1.00}^{+2.17}$ & $16.70$ & $16.47$ & $17.83_{-1.03}^{+2.27}$ & $17.07$ & $17.49$ \\
$\chi^2_{\rm HIP}$ & \nodata &  \nodata & \nodata & \nodata & \nodata & \nodata & \nodata & $108.4_{-13.6}^{+14.8}$ & $111.1$ & $110.2$ \\
$\chi^2_{\rm Gaia-Pos.}$ & \nodata &  \nodata & \nodata & \nodata & \nodata & \nodata & \nodata & $1.84_{-1.20}^{+1.85}$ & $0.10$ & $0.01$ \\
$\chi^2_{\rm Gaia-PM}$ & \nodata &  \nodata & \nodata & \nodata & \nodata & \nodata & \nodata & $1.40_{-1.05}^{+2.30}$ & $0.01$ & $0.08$ \\
$\chi^2_{\nu}$ & \nodata & $1.15_{-0.11}^{+0.17}$ & $0.91$ & $0.95$ & $1.32_{-0.13}^{+0.17}$ & $0.94$ & $1.11$ & $1.11_{-0.11}^{+0.12}$ & $1.06$ & $1.06$ \\
\enddata
\tablecomments{Orbits with $\Omega < 180$\,deg were wrapped by $\omega+180$, $\Omega+180$, $\tau$ in fractions of the orbital period since MJD 57412.1335. Reduced $\chi^2$ calculated using 22 degrees of freedom for the visual orbit fit, and 125 for the combined fit.}
\end{deluxetable*}

\section{Disk Modeling}
\label{sec:disk_modeling}

\subsection{Modeling Setup}


We model the submillimeter emission from the disk with MCFOST, a radiative transfer code for circumstellar modeling \citep{Pinte2006}. In short, the system is set up as a central star surrounded by a circumstellar disk where the stellar radiation is propagated through the disk using a Monte Carlo process to evaluate the dust temperature in all locations, and the resulting thermal emission map is generated using a ray tracing method. We assume that the disk is axisymmetric, and that its dust population can be represented by a uniform composition \citep[astronomical silicates from][]{Draine1984} and power law size distribution. We assume an $N(a) \propto a^{-3.5}$ distribution \citep{Dohnanyi1969} that extends from 3\,$\mu$m, the approximate blowout size for a mid-F star \citep[e.g.,][]{Pawellek2015}, to 1\,mm. As is applicable for optically thin disks, the dust grains are assumed to be in radiative equilibrium with the stellar radiation field but local thermal equilibrium is not enforced. This leads to the smaller dust grains being super-heated compared to large grains and to the blackbody approximation. The star emission is simulated using a 6600\,K stellar spectrum generated by the PHOENIX grid with a total luminosity of 3.1\,$L_\odot$ \citep[see eg.,][]{Moor:2015kf}. 

Following standard practice for debris disks \citep[e.g.,][]{Augereau1999A,Esposito2020}, we select the following prescription for the disk density as a function of position between the disk's inner($R_{in}$) and outer radius($R_{out}$):
\onecolumngrid 
\begin{gather}
\label{density}
\rho(r,z) \propto \left(\left(\frac{r}{R_c}\right)^{-2p_{in}} + \left(\frac{r}{R_c}\right)^{-2p_{out}}\right)^{-1/2 } \times \text{exp}\left(-\left(\frac{|z|}{h(r)}\right)^{\gamma_{\text{vert}}}\right).
\end{gather}
\twocolumngrid
We fix $\gamma_{\text{vert}}=1$ for an exponential vertical profile. The corresponding scale height, $h(r)$, is assumed to be a linear function of radius, i.e., the disk has a bow-tie shape. In this case, $h = h_0 \frac{r}{r_0}$ where $r_0$ is an arbitrary reference radius. We selected $r_0 = 100$\,au. The surface density profile is a smoothly connected pair of power law regimes, with exponents $p_{in}$ and $p_{out}$ at radii $r \ll R_c$ and $r \gg R_c$ if $p_{in} > 0$ and $p_{out} < 0$, as is usually the case. The peak surface density occurs near, but not exactly at, $r = R_c$, depending on the value of $p_{in}$ and $p_{out}$ \citep{Augereau1999A}. The total dust mass, $M_d$, is obtained by integrating Eq.\,1 from the disk's inner and outer radii, $R_{\text{in}}$ and $R_{\text{out}}$, respectively.

The disk image is then produced for a combination of inclination ($i$) and position angle (PA) with a pixel scale of 0\farcs2/pixel, convolved with a 2D Gaussian beam constructed with the major and minor axes and the position angle from the ALMA beam. To reduce the issue of correlated noise in the interferometric map, we rebin the observed and model images to a 1\arcsec/pixel scale so that each pixel can be reasonably considered as independent of its neighbors. The two images are aligned based on the 2D Gaussian fit to the observed image and a $\chi^2$ goodness-of-fit metric is computed based on a 20\arcsec$\times$20\arcsec\ field of view centered on the disk.

With this routine established, we proceed to perform a Monte Carlo Markov Chain (MCMC) run to determine the best fit parameter for the disk. We vary a total of 9 parameters: $i$, PA, $M_d$, $R_c$, $R_{\text{in}}$, $R_{\text{out}}$, $p_{\text{in}}$, $p_{\text{out}}$, $h_0$. The values of the priors are given in Table~\ref{tbl:ALMA}.  The ensemble initializes the walkers with flat priors for every parameter except for the dust mass, which was sampled from a log uniform prior.  
With these priors, we set up the two-temperature MCMC with 100 walkers both for the hot and cold chains. We ran the chain for 2000 iterations with a script adapted from {\sl diskmc} \citep{Esposito2018} and verified that the chain had reached convergence by the end of the run. The first 600 iterations were rejected as burn-in steps. 

\subsection{Results}



   
   \begin{deluxetable*}{cccccc}
\tablewidth{0pt}
\tablecaption{Disk physical parameters from ALMA map fitting\label{tbl:ALMA}}
\tablehead{
\colhead{Parameter} & \colhead{Unit} & \colhead{Prior Range} & \colhead{Best Fit-ALMA} & \colhead{Confidence Interval}&\colhead{Best Fit-SED}}
\startdata
	$M_d$ & $M_{\oplus}$ & ($4.2\times\,10^{-5}$, $3.3\times\,10^5$) & 0.0182 & $0.0192^{+0.0016}_{-0.0015}$ & 0.0158 \\
	PA & \degr & (-60.,120.) & 32.7 & $32.7^{+0.7}_{-0.8}$ & 33.6 \\
	$i$ & \degr & (0., 90.) & 84.7 & $\geq 79.3$ & 87.1 \\
	$p_{\text{out}}$ & & (-10., 0.) & -1.2 & $-1.2^{+0.3}_{-0.4}$ & -0.8\\
	$p_{\text{in}}$&  & (0,10.) & 3.05 & $\leq 10.2$ & 7.6\\
	$R_c$ & au & (20.,150.),&41.0 & $\leq 114.4$ & 47.6 \\
	$R_{\text{in}}$ & au & (0.1, 75.)&20.3 & $\leq 62.3$ & 34.0 \\
	$R_{\text{out}}$ & au & (200.,401.) & 260.0 &$258.8^{+26.8}_{-21.2}$  &234.3 \\
	$h_0$& au & (0.5., 20.) & 1.58 & $\leq11.0$ & 3.91\\
\enddata
\tablecomments{ We report the prior ranges, best fit image parameter, the 1$\sigma$ confidence interval and the best fit SED parameter. For less constrained parameters, 3$\sigma$ upper or lower limits are reported instead. The reference radius is set to $r_0=100$\,au.}
\end{deluxetable*}

\begin{figure*}
   \centering
   \includegraphics[width=\textwidth]{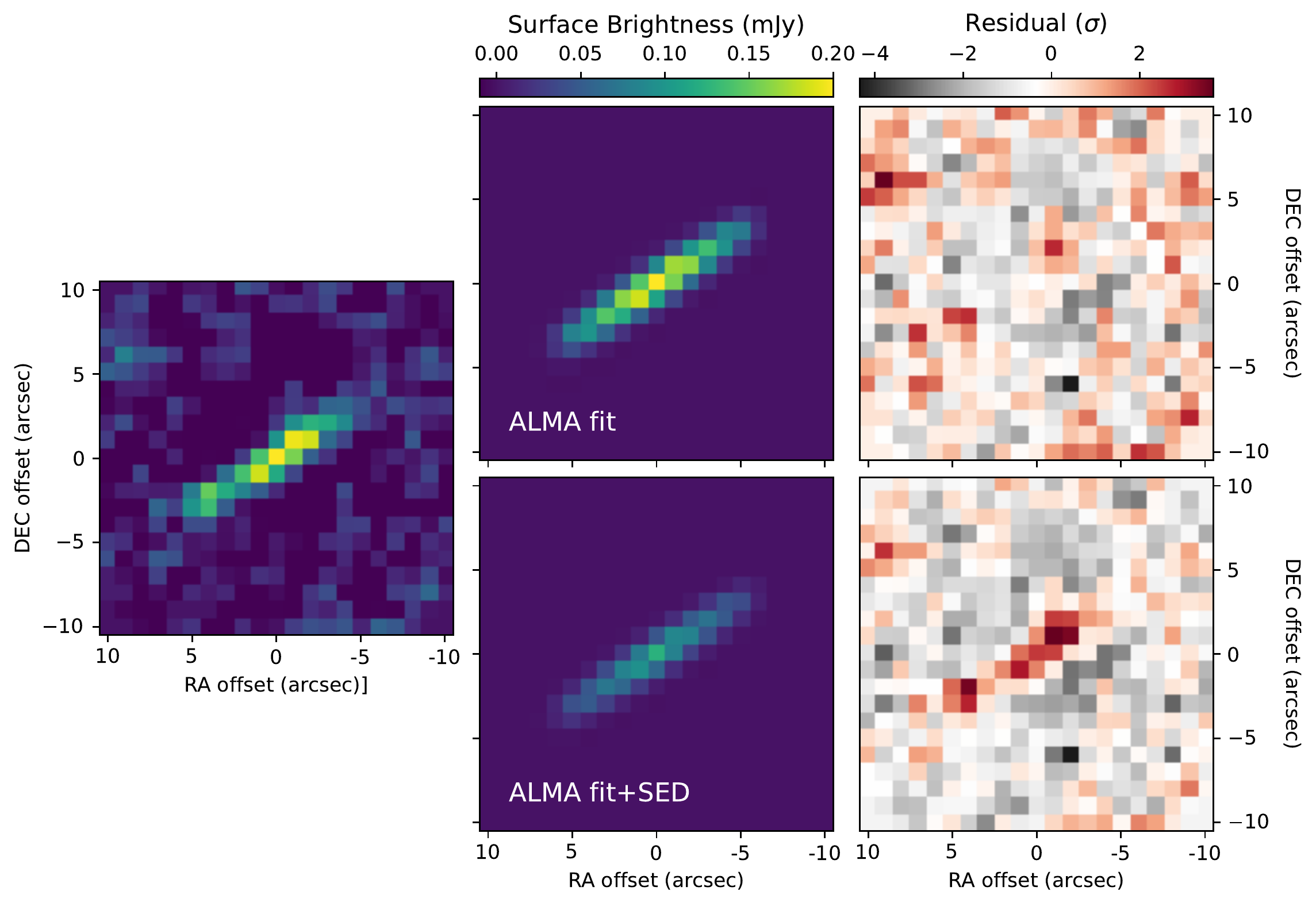} 
   \caption{From left to right: ALMA map of the system binned to 1\arcsec\ pixel; best fit model to the ALMA data (top) and the model from the converged portion of the chain that best matches the SED (bottom, see Section\,5.2); residuals for both models.  The ALMA maps uses the same colorbar as the models. The ALMA fit agrees well with the model image while the SED-selected model has an acceptable geometry but is significantly dimmer.  \label{fig:best_fit}}
\end{figure*}

\begin{figure*}
   \centering
   \includegraphics[width=\textwidth]{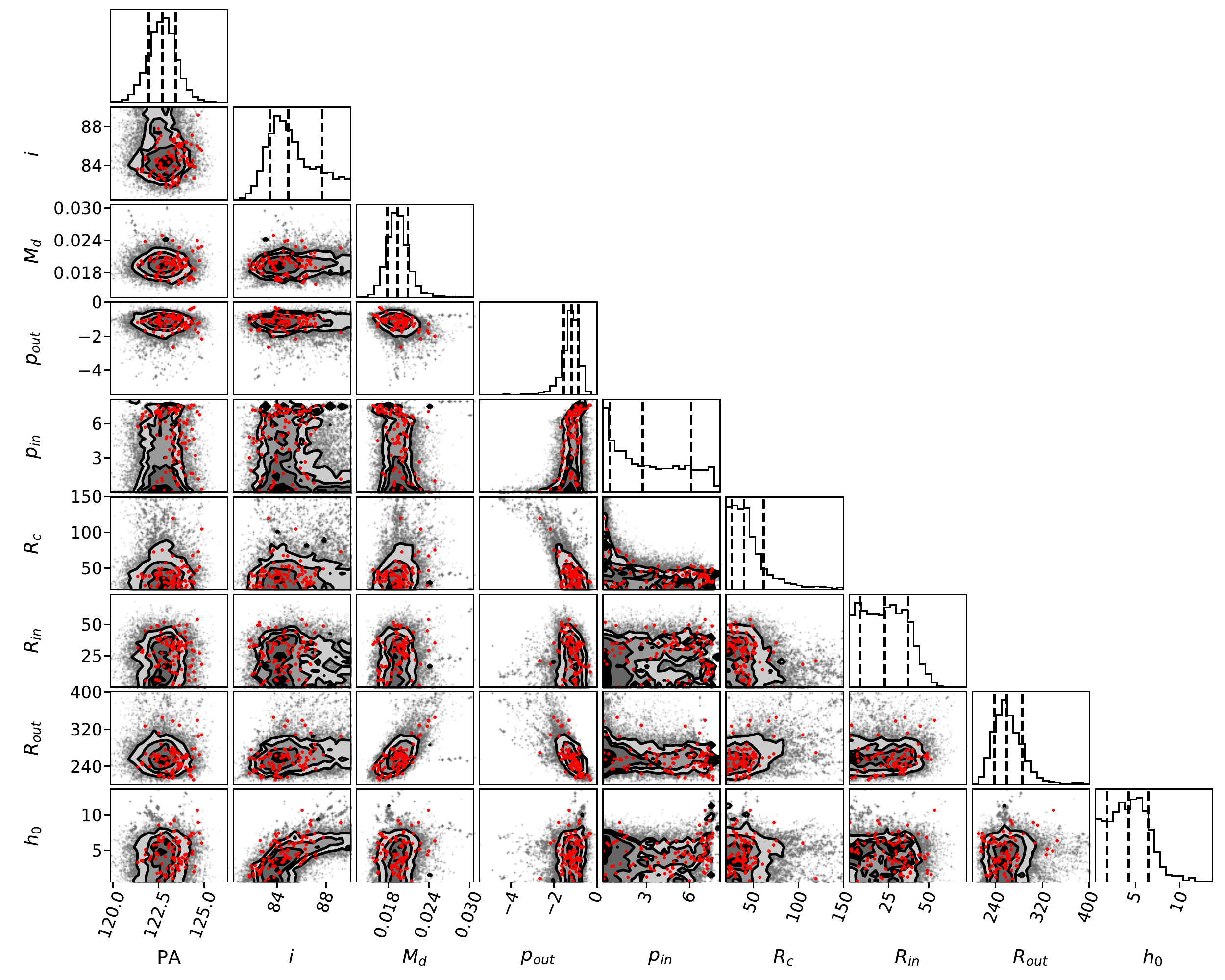}
   \caption{Results of the MCMC fit of the ALMA image of the HR~2562 disk properties. The cornerplot, shown as grey dots and black density contours, is created from the last 1400 iterations of the chain, i.e., after convergence was reached.  Contours represent the 1, 2, and 3\,$\sigma$ confidence levels. The 100 models selected from the converging part of the chain that best fit the observed SED of the disk (see Section\,5.2) are represented in red dots.
   \label{fig:corner_overplot}}
\end{figure*}

The best fit model and the associated residuals are presented in the top row of Figure \ref{fig:best_fit}. This model describes the data reasonably well ($\chi_{red}^2 = 1.50$) with nearly random residuals and, especially, no structured residuals at the location of the disk emission. Considering the entirety of the MCMC chain (Figure \ref{fig:corner_overplot}), we find that several parameters (PA, $M_d$, $R_{\text{out}}$ and $p_{\text{out}}$) are well constrained and we report their $1\sigma$ confidence intervals in Table~\ref{tbl:ALMA}. 
For the remainder of the parameters 
we obtained 3$\sigma$ upper ($R_{\text{in}}$, $p_{\text{in}}$, $R_c$ and $h_0$) or lower ($i$) limits from our posteriors.
As expected given that the disk is well resolved in our ALMA observations, several of the geometric parameters are well constrained. In addition, the disk mass is well constrained by the integrated flux of the disk.

The corner plot reveals important correlations between parameters. First of all, there is some ambiguity between the disk inclination and its vertical extent. Specifically, the ALMA map is consistent with a vertically thin disk inclined at about 84\degr,but a thicker disk almost exactly edge-on is also consistent with the observations. This is unsurprising given the $\approx35$\,au linear resolution of these observations; higher resolution mapping will be necessary to tighten the constraints on these parameters. 
The second correlation connects the disk radial extent and its surface density profile. For instance, higher values of $R_c$ are associated to a steeper outer density profile. Furthermore, the inner disk radius and inner density profile slope have broad ranges of allowed value since the resolution of our ALMA observations does not allow us to fully resolve the disk inner hole. We place a 3$\sigma$ upper limit on both parameters from the ALMA map and will further explore their relations with other available information in the next section.



\section{Analysis}
\label{sec:analyis}
\subsection{System geometry}
\label{sec:system geometry}
Since the discovery of the HR~2562 B substellar companion and the initial mapping of the disk structure with {\it Herschel}, the observations presented here allow us
to further our understanding of the system: the continued imaging of the companion provides tighter constraints on its orbit, the Gaia DR3 releases opens the door to a dynamical mass estimate on the brown dwarf via the absolute astrometry of the star, and the better resolved ALMA data further clarifies the geometric structure of the disk. Most importantly, with the updated companion orbit and the resolved image of the disk, we are in a good position to quantify the system geometry and obtain concrete evidence of direct interaction between HR2562B and the debris disk. 

\begin{figure*}
   \centering
   \includegraphics[width=\textwidth]{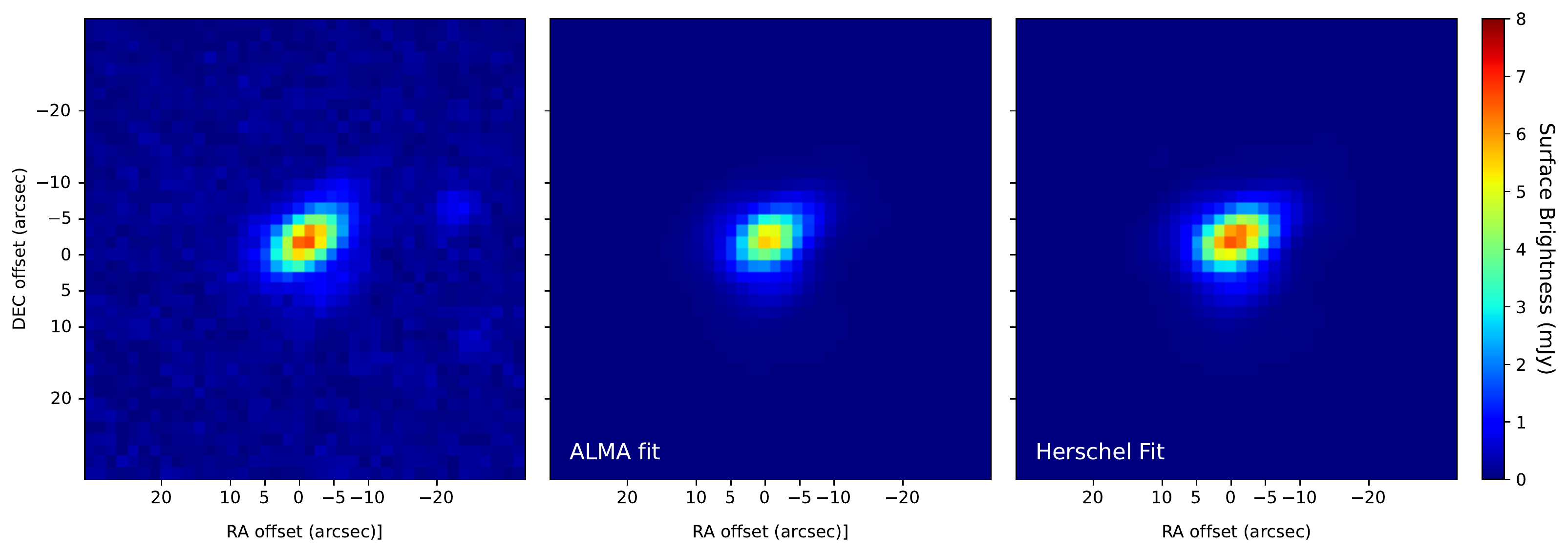} 
   \caption{Observed \citep[left panel,][]{Moor:2015kf} and synthetic (center and right panels) 70\,$\mu$m Herschel image of HR~2562. The two models represent the model using the ALMA best fit parameters from this study and the {\it Herschel} best fit model from Moor et al.}
   \label{fig:Herschel_ALMA}
\end{figure*}

We first focus on our updated constraints on the disk properties. A similar analysis was performed in \cite{Moor:2015kf} to derive the properties of the disk. 
The acceptable ranges for the geometrical disk parameters are consistent between their study and our fit result, although $R_{\text{in}}$ and $R_{\text{out}}$ are less constrained by the {\it Herschel} observations due to their lower resolution, which leads to the disk being only marginally resolved. To further contrast the ALMA observation with the {\it Herschel} observation, we recreate the Herschel best fit model using MCFOST with the single power law density profile and the best fit parameters from \cite{Moor:2015kf}, and adjusting the dust mass so that the total flux of the disk model matches the measured SED of the system at 70$\mu$m. The resulting image is then convolved with the PACS beam. We also produce the 70 $\mu$m image of the best fit model to the ALMA data, again convolving it with the PACS beam. The comparison between observed and synthetic Herschel observations is presented in Figure\,\ref{fig:Herschel_ALMA}. The Herschel disk looks geometrically similar to our best fit result as we expected, and we observe an overall weaker integrated brightness in the Herschel model compared to the ALMA model. The derived FWHM of the two models are consistent with each other with the two models both having a weaker integrated brightness than the observation. Overall, the best fitting model for the ALMA map is also consistent with the Herschel image at 70 micron.

\begin{figure}
   \centering
   \includegraphics[width=0.5\textwidth]{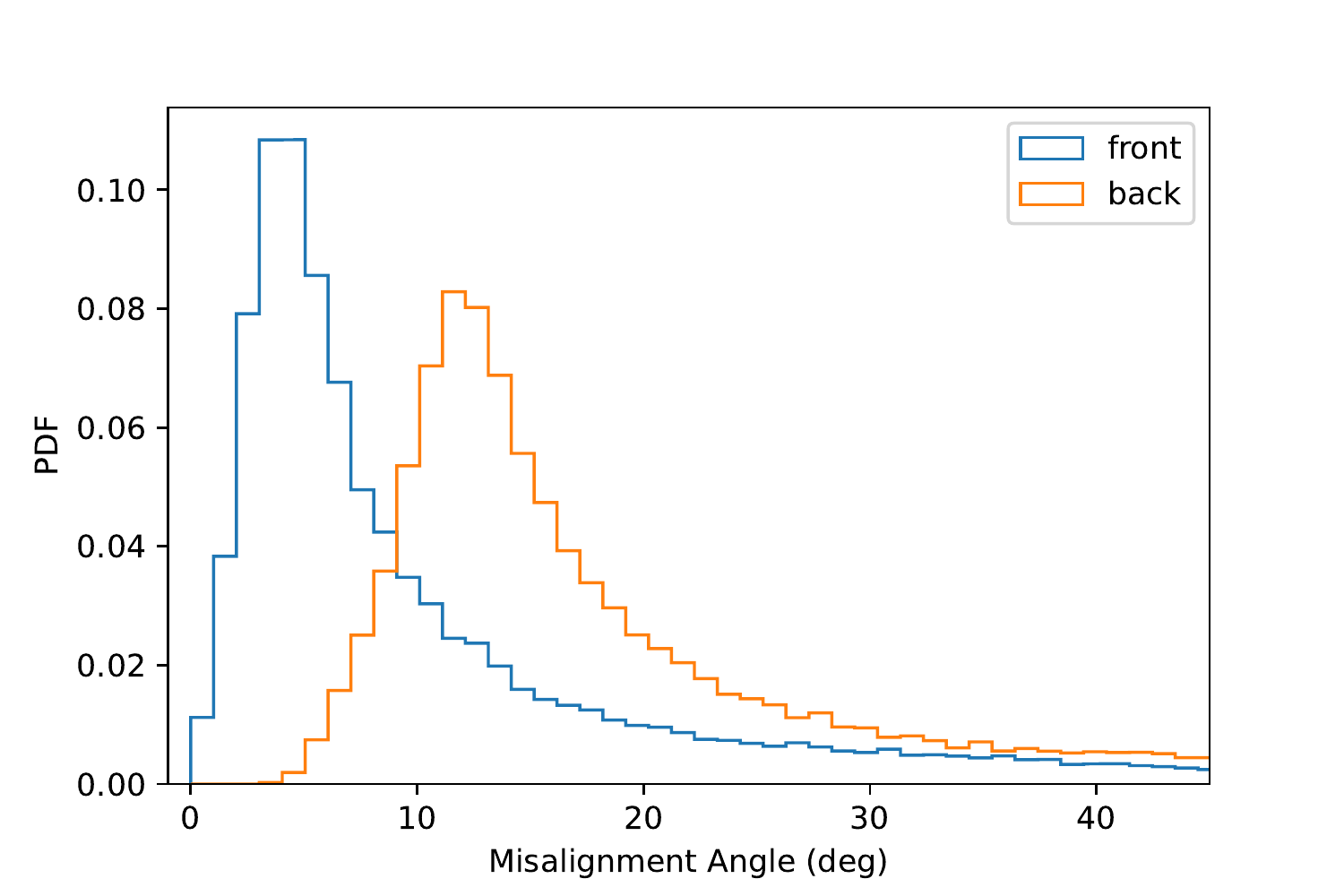} 
   \caption{Relative inclination of the disk (as seen by ALMA) and the companion's orbit. The two histograms represent the two cases regarding the front/back ambiguity in both disk and orbital planes.}
   \label{fig:misalignment}
\end{figure}

Combining the updated orbit fit and disk analysis, we can study the alignment between the disk and the plane of the orbit of the companion. To do this, we calculate the misalignment angle with equation 1  in \cite{Czekala2019} and the unconstrained relative astrometry fit from section 3.2.1. The result is shown in Figure \ref{fig:misalignment}. 
The blue and yellow lines represent the front and back ambiguity between the disk and the orbit. This is because it is impossible to disambiguate between the orbital angular momentum vector pointing out of, or into, the plane of the sky. 
In either case, we find that the system is likely in a near-coplanar situation. 
Specifically, in the ``front" case, where both angular momentum vectors points the same way relative to the plane of the sky, we report a 1 $\sigma$ confidence interval of $7.0^{+ 17.1}_{-3.7}\,\degr$. In the other case, we report a 1 $\sigma$ confidence interval of $15.2^{+ 17.7}_{-4.8}\,\degr$. The low-probability tail of significant misalignment in both cases can be traced back to the uncertainty in the companion's orbit. We further note that the more misaligned solutions also have higher eccentricity, which we consider to be less likely.  
Continuous astrometric monitoring will make this upper limit much more stringent if the disk and orbit are indeed close to coplanar.

\begin{figure*}
\includegraphics[width=\textwidth]{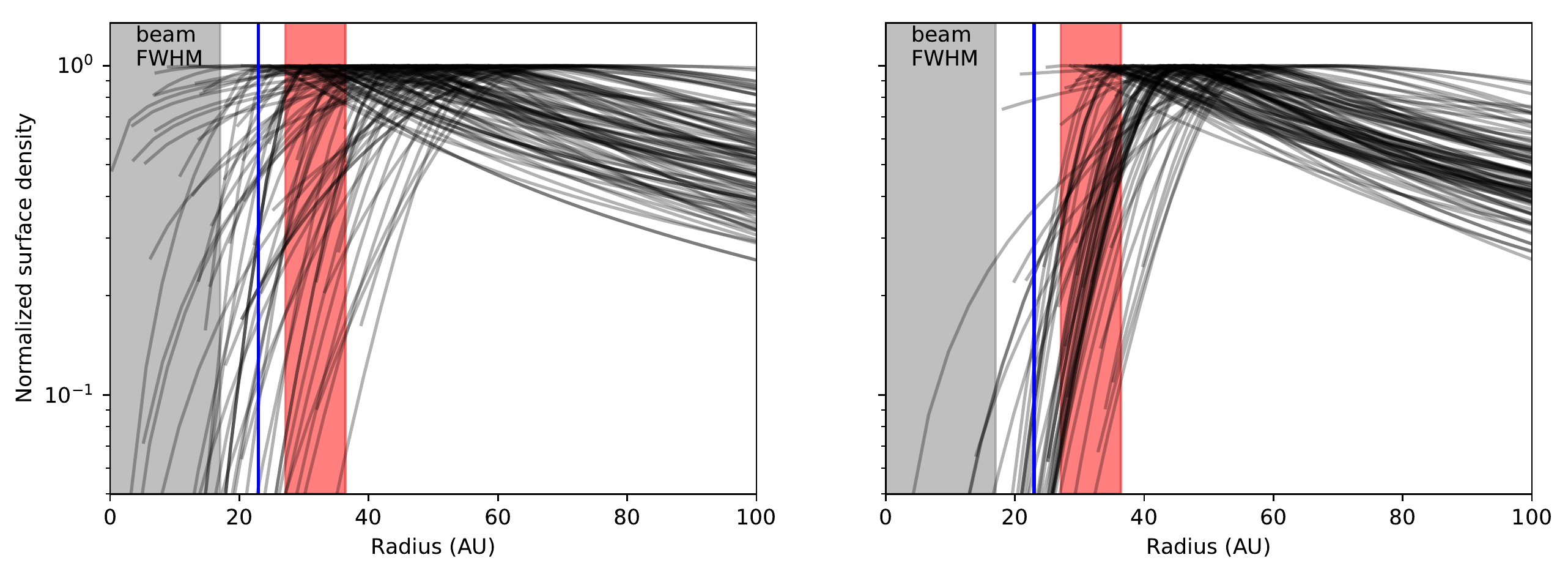} 
\caption{Left: density profiles of 100 randomly selected models from the converged part of the MCMC chain; right: density profiles of SED-selected models (see Section \ref{sec:bestfit SED}). The density profiles in both panels are normalized to their maximum value for ease of comparison. The blue line marks the separation of the brown dwarf at the time of the ALMA observations while the red bar represent the 1\,$\sigma$ range of apoastron distance of its orbit. The restriction of SED alone is enough to eliminate most models with $R_{in}$ smaller than the current separation of the brown dwarf, which are physically implausible.\label{fig:Density_random}}
\end{figure*}

We continue to explore the density structure of the disk from our findings. The density profiles of 100 fitting models selected from the converged part of the MCMC chain are plotted in the left panel of Figure \ref{fig:Density_random}. Interestingly, a number of models that satisfyingly fit the ALMA map have density profiles that extend down (even interior) to the current separation of the brown dwarf companion. This is physically implausible, as we expect the massive companion to clear out dust out to at least its apoastron distance. However, the fit to the ALMA map was not informed by the location and orbit of the companion, so it is reassuring to note that at least some of the models have peak surface density radii that are exterior to the companion's orbit. Indeed, many of these models have significant amounts of dust close to the apoastron distance, suggesting that the brown dwarf may be 
directly interacting with it. Our SED selection in the following subsection provides additional support to this conclusion. 

\subsection{Best fitting SED}
\label{sec:bestfit SED}
So far we have only considered the ALMA disk image in our analysis, but a holistic approach needs to consider multiple types of observation. An important piece of information is the spectral energy distribution (SED) of the disk. In their {\it Herschel}-based study, \cite{Moor:2015kf} fit the SED and the image of the disk separately. The SED fit in \cite{Moor:2015kf} yields a radius of $64\pm6$ au, which is about twice as large as the image fit (and also consistent with our fit results). A likely explanation for this discrepancy is the assumption in the SED fit that the dust is at blackbody equilibrium temperature. In debris disks, the smallest grains tend to super-heat to a temperature that is significantly hotter. As a result, SED-informed disk radii are significantly smaller than those obtained from resolved imaging.
This discrepancy has also been studied quantitatively with previous {\it Herschel} observations of entire samples of debris disks and a clear trend with stellar luminosity has been shown \citep[e.g.,][]{Morales2016}. Given the luminosity of HR~2562, we conclude that a factor of 2 discrepancy between SED- and image-fit disk radii is consistent with previous literature.

\begin{figure*}
\includegraphics[width=\textwidth]{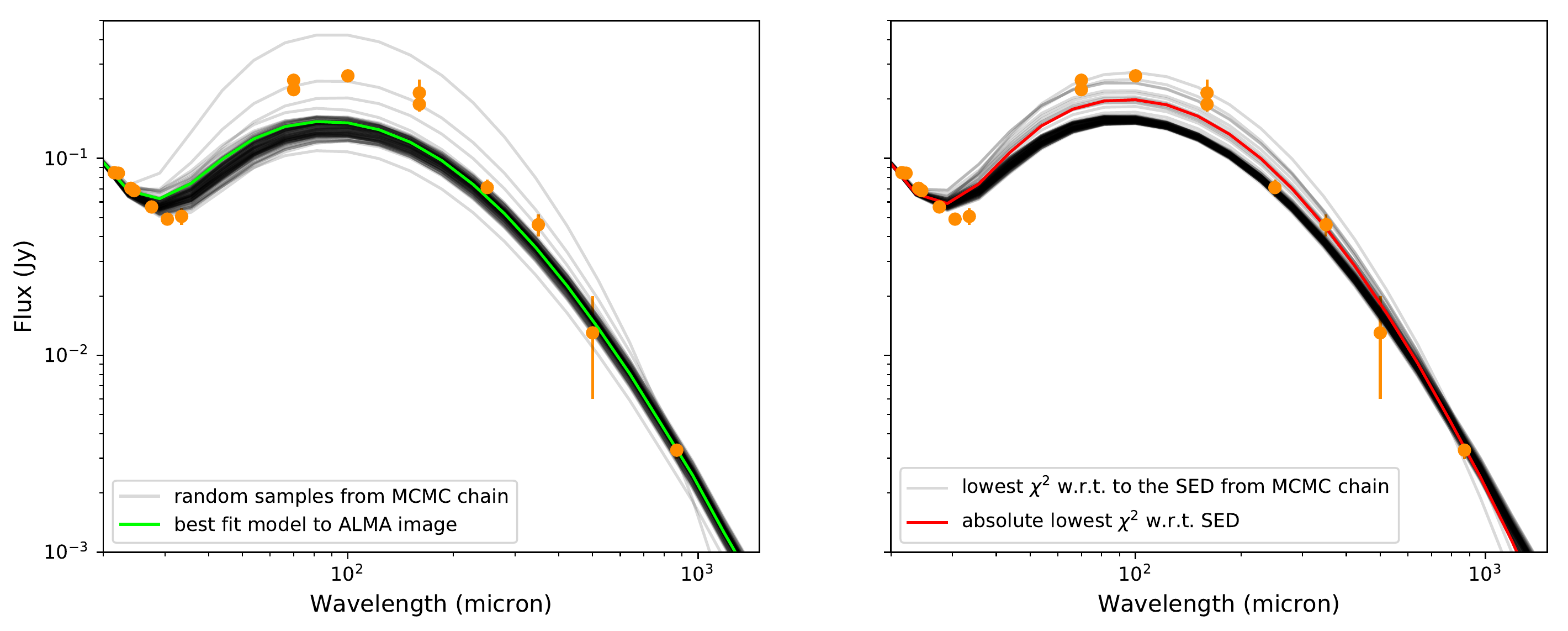} 
\caption{SED of HR~2562 (orange circles) compared to subsets of the models from the converged part of the ALMA map fitting MCMC chain. The left panel shows 100 models randomly selected from the chain without considering the SED while the right panel shows the 100 models  from the same portion of the chain with the least $\chi^2$ with respect to the SED. The green solid line in the left panel represents the model with the best fit model to the ALMA image, and the red solid line in the right panel represents the lowest $\chi^2$ with respect to the measured SED. \label{fig:sed}}
\end{figure*}

Nonetheless, our radiative transfer model, which self-consistently treats the cooling inefficiency of small dust grains, also predicts a system's SED so that we can in principle incorporate this observable in our analysis. This allows us to consider the SED of the models in the MCMC chain to assess whether more stringent constraints on the disk properties can be inferred. Unfortunately, we did not compute the SED during the chain and this analysis must be conducted a posteriori.
To examine this, we computed the SEDs of a selection of models that were included in the MCMC chain. Specifically, we randomly selected 1000 models from all the walkers in the  lower temperature chain between steps 1900 to 2000, where the chain was well converged. We then computed the $\chi^2$ between the models and the observed SED (including the IRAS, {\it Herschel} and ALMA fluxes) and selected the 100 best fit models. To put the result of this SED selection in context, we also randomly selected 100 models from the same part of the chains and plot their SEDs. The results are shown in Figure \ref{fig:sed}. Many of the models selected based exclusively on the ALMA image fit are poor fits to the SED, generally underestimating the 70--160\,$\mu$m emission. Nonetheless, some models are good fits to the SED even though our model fitting was only informed by the ALMA image.

To gauge how well the best fit SED model describes the ALMA image, we compare it
with the best fit image model and the data in Figure \ref{fig:corner_overplot}. The best fit SED model does not fit the data as well, with a \textbf{reduced }$\chi^2$ of 1.68 and the clump-like structure around the location of the star in the disk, indicating that the SED-fitting model produces a less centrally peaked surface brightness profile. To see the effect of SED rejection on the disk structures, we overplot the models selected by lower SED $\chi^2$ with red dots in  Figure \ref{fig:corner_overplot}. The main difference between the SED-selected models and the overall MCMC posteriors is that the former 
show a preference for larger values of $R_{\text{in}}$. To illustrate this, we again compare the density profiles of 100 randomly selected models with those of the 100 best fit SED models as seen in the right panel of  Figure \ref{fig:Density_random}. 
The best fit SED models almost universally have a peak surface density radius that is exterior to the apoastron distance of the companion.
In other words, the best SED models prefer the family of models with an inner hole potentially consistent with truncation by the brown dwarf. This shows that the SED indeed provides useful information for us to narrow down the result from the ALMA image MCMC. 

Overall, although our SED-selected models 
do not fit the ALMA image as well, they are consistent with strong dynamical interaction between the substellar companion and the debris disk, providing a new testbed for such interactions. Future work will require a simultaneous fit to the SED and resolved images of the disk to provide tighter and more consistent constraints of the exact architecture of the system.

\subsection{Dynamical Interaction}

A number of theoretical and numerical studies have tackled the problem of disk truncation by an interior substellar object, but none was tailored to the exact configuration of the HR~2562 system. The companion mass, for which we only have a dynamical upper limit and a model-dependent estimate, the companion semimajor axis and the orbital eccentricity, which is only modestly constrained at this point, are three main factors that set the location of the disk's inner radius.
Here we will explore the predictions of several such models 
when applied to the parameter values estimated in this study. Relative inclination likely also plays a role, although most studies assume (near) coplanarity. Since this is assumption is consistent with all the data at hand for HR~2562, we can thus use these theoretical predictions to compare them to the geometry of the HR~2562 system and assess the likelihood that the brown dwarf companion is actively truncating the debris disk. It is worth pointing out that disk truncation studies sometimes rely on different criteria to define the inner edge, such as the ``chaotic zone" \citep{Quillen2006,Chiang2009} or the ``Hill sphere argument" \citep{Pearce2014}, so that for a given set of parameters, there is a range of predicted inner radii for the truncated disk. Nonetheless, we can compare the range of predicted inner disk radii to the range derived from fitting our ALMA observations. Although there is a possibility that the companion's orbit may be highly eccentric, which may produce the clump we see in Figure 3 as well as an asymmetric disk,  we do not think this is the likely scenario due to the clump still being marginally consistent with a symmetric disk in brightness profile as well as the fact that the disk and the orbit are shown to align with each other (Figure \ref{fig:misalignment}). Thus, we only consider the case for low to moderate eccentricity orbits in the following analysis.

\begin{figure}
   \centering
   \includegraphics[width=0.5\textwidth]{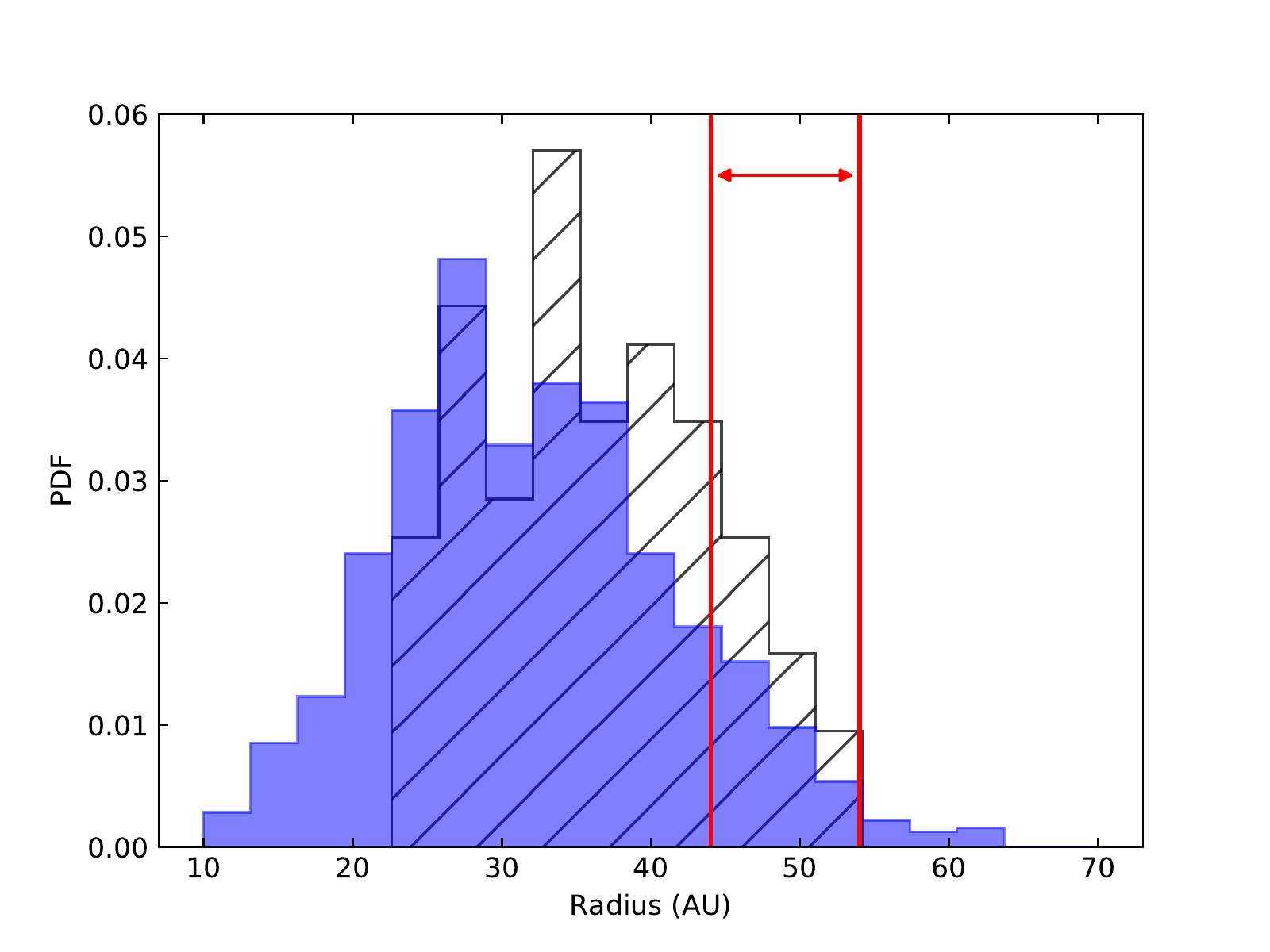} 
   \caption{Distribution of inner disk radius for the 1000 randomly selected models from the converged part of the ALMA MCMC fit and the 100 best fitting models by SED are plotted (solid and hatched histogram, respectively). The red arrow and lines mark the range of predicted disk inner edge locations using the dynamical models from \cite{Chiang2009}, \cite{Pearce2014} and \cite{Rodigas2014}. 
   \label{fig:Rin_estimate_hist}}
\end{figure}

Given our limited constraint on the companion's orbit, we select three different values of eccentricity and derive the corresponding disk inner edge location. We obtain the corresponding semimajor axis from the orbit posterior distribution in Figure \ref{fig:orbit-corner2} and conservatively assume the upper limit of 18.5 M$_\text{Jup}$ as  the companion mass in our predictions.We first use an eccentricity of $e\sim0.1$ (and a semi-major axis of 28\,au) and adopt equation\, 16 from the Fomalhaut-tailored simulations of \cite{Chiang2009} that is designed for low eccentricity systems. This yields an estimate of the disk's inner radius of 44\,au. We also considered the analysis of \cite{Pearce2014} and applied their Equations\,9 and 10, assuming a semi-major axis of 
28\,au and an eccentricity of $e\sim 0.1$, to predict an inner radius of 54\, au. Finally, we consider the result from the N-body disk simulations in \cite{Rodigas2014} for a more general framework using the same methodology, exploring explicitly a broader range of perturber mass and orbital eccentricity (up to $e=0.2$). From their study, we adopt the 10\,$M_J$ case as it is the closest to the likely mass of HR~2562~B. Taking results from their Table 2 and adopting semi-major axes of  30\,au and  24\,au for the circular and $e=0.2$ cases, we obtain a predicted location of the disk inner radius of  50.8 and 51.1\,au, respectively. To compare the estimates with our models, we define the effective inner radius of our models as the half peak point of their surface density. We compute the effective inner radii for 1000 models from the converged portion of the ALMA MCMC chain as well as the 100 best fitting models by SED, and show the results alongside the above estimates on Figure \ref{fig:Rin_estimate_hist}.

All in all, existing dynamical models predict that the HR~2562~B mass and orbit should result in the exterior debris disk being truncated at about 45--55\,au. Our ALMA observations suggest that disk extends at least that close to the brown dwarf, possibly even closer (see  Figure \ref{fig:Rin_estimate_hist}). It is therefore extremely likely that the brown dwarf is directly responsible for the disk truncation, without the need for additional perturbers, as has been suggested for HD~206893. Given the limited orbital coverage and modest resolution of our ALMA observations, it is currently impossible to discriminate between the different models discussed above, but it is likely that further observations in the next few years can produce observational constraints that allow for such a study. HR~2562 is therefore bound to become a new testbed for dynamical studies of debris disk truncation by an interior substellar companion.

\section{Conclusion}
\label{sec:conclusion}
HR~2562 is an exciting system with a brown dwarf companion located inside the inner hole of a cold debris disk, making it a very likely candidate for dynamical sculpting between the disk and the companion. In this paper we report the first ALMA 870\,$\mu$m observations of the disk around HR~2562, as well as the continued GPI monitoring on the substellar companion's orbital motion, which is close to edge-on. Coupled with Gaia EDR3 observations of the star we obtain improved constraints on the substellar companion orbital elements. 
In particular, the combined analysis of the absolute and relative astrometry of the system places a $3\sigma$ upper limit of 18.5\,$M_{\text{jup}}$ on the companion. 

The new ALMA observations with an angular resolution roughly 6 times higher than previous Herschel observations of the system achieved a peak signal to noise ratio of $\sim 7$, enabling a more detailed study of the disk structure. Using radiative transfer modeling, we perform an MCMC fit of the disk key parameters and confirm the disk is see at high inclination ($3\sigma$ lower limit of $i >79$\fdg3), consistent with previous studies of the system. We further compute the misalignment angle between the disk and the orbit and find that it is either $7^{+17}_{-4}$\,\degr\ or $15^{+18}_{-5}$\,\degr, depending on unresolved ambiguities due to projection effects.
This provides further evidence that the disk and the orbit are close to a coplanar configuration. 

To test dynamical models of disk truncation by a low-mass companion, we focus on the location of the inner edge of the debris disk. Modelling of the ALMA map yields a $3\sigma$ upper limit of 62.3\,au, as the resolution is insufficient to fully resolve the inner parts of the disk. We note, however, that consideration of the SED in the analysis further narrows the allowable range of inner disk radii and locate the latter at $\approx 30$\,au, albeit with a significant uncertainty. This is close to the apoastron distance derived from our orbital fit, providing further evidence for the companion dynamically sculpting the disk.
We further test three disk truncation models, which predict the location of the disk inner radius to be in the 45--55\,au range based on the companion's orbit and estimated mass. The image-derived inner radius is even closer to the companion, possibly pointing to a lower mass estimate for the companion and/or to shortcoming in the models. 

HR~2562 presents a unique opportunity to quantitatively test disk truncation models, provided the system's architecture can be further constrained. Continued monitoring of the companion's orbit with high-contrast imaging instruments and of the reflex motion of the host star with Gaia will yield increasingly precise estimates of the orbit geometry as well as the first dynamical measurement of the companion's mass. Deeper and higher-resolution images with ALMA, as well as a simultaneous fit to the sub-millimeter map and the system's SED, will provided a better defined view of the disk's inner regions, directly testing dynamical models.

\acknowledgments

This work is based on observations obtained at the Gemini Observatory, which is operated by the Association of Universities for  Research in Astronomy, Inc. (AURA), under a cooperative agreement with the National Science Foundation (NSF) on behalf of the Gemini partnership: the NSF (United States), the National Research Council (Canada), CONICYT (Chile), Ministerio de Ciencia, Tecnolog\'{\i}a e Innovaci\'on Productiva (Argentina), and Minist\'erio da Ci\^encia, Tecnologia e Inova\c{c}\~ao (Brazil). This work made use of data from the European Space Agency mission Gaia (https://www.cosmos.esa.int/gaia), processed by the Gaia Data Processing and Analysis Consortium (DPAC, https://www.cosmos.esa.int/web/gaia/dpac/consortium). Funding for the DPAC has been provided by national institutions, in particular the institutions participating in the Gaia Multilateral Agreement. This research made use of the SIMBAD and VizieR databases, operated at CDS, Strasbourg, France. We thank support from NSF AST-1518332, NASA NNX15AC89G,  NNX15AD95G/NEXSS, and award SOSPA4-006 through NSF from the NRAO.

\appendix


\section{GAIA goodness of fit metric}
\label{append:gaia}
 
\begin{figure}[]
\plotone{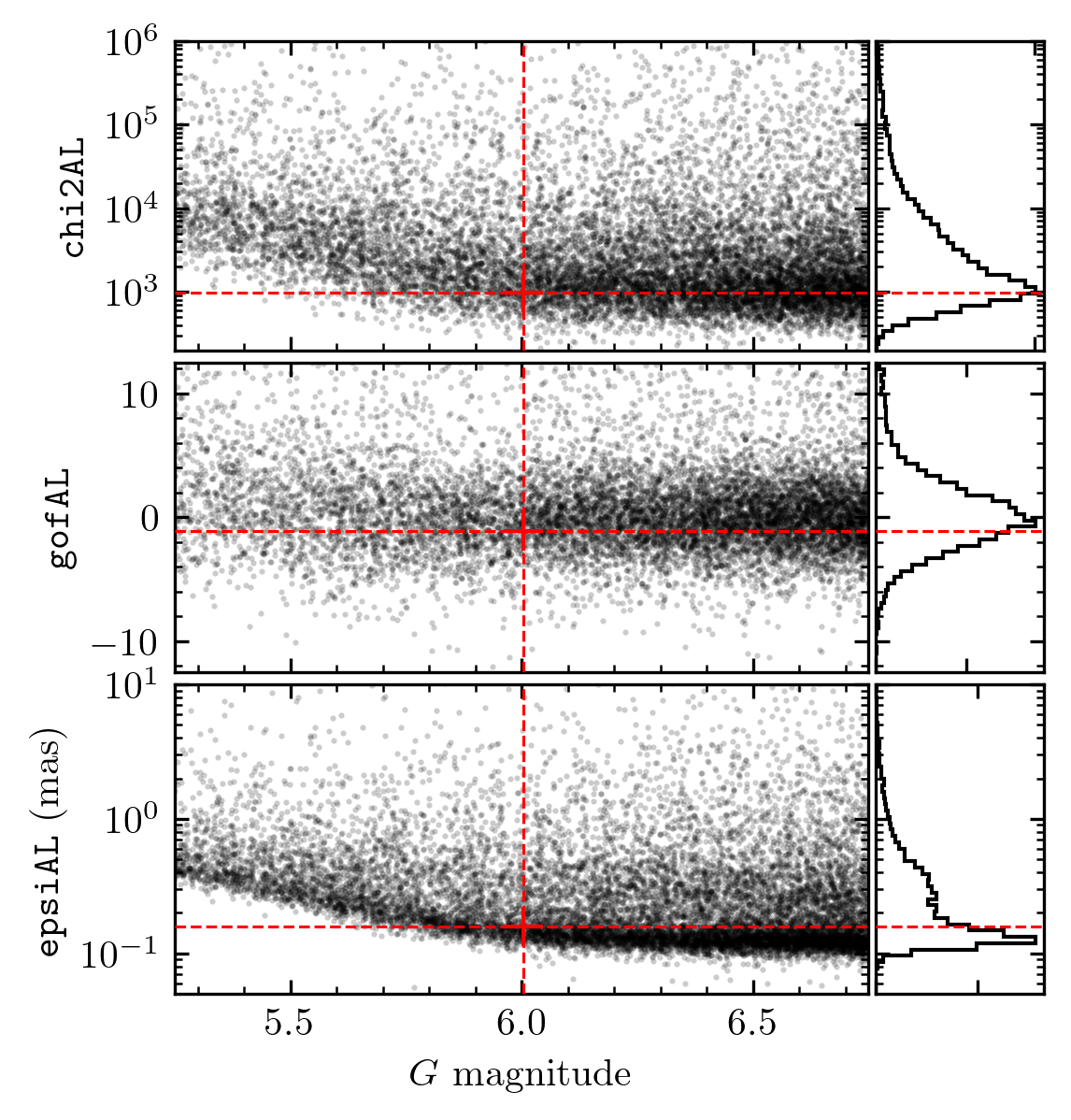} 
\caption{Goodness of fit metrics for HR~2562 (red cross) compared with other {\it Gaia} sources of a similar $G$ magnitude (\textbf{dark points}). The top two panels give the $\chi^2$ and a goodness of fit metric, while the bottom panel gives the amplitude of the residuals to a five-parameter fit that assumes linear space motion. The distribution of these three parameters are shown on the right. Large values for any of these parameters can be used as an indication of non-linear motion between mid-2015 and mid-2017 caused by an orbiting companion.\label{fig:gaia-stats}}
\end{figure}

The GAIA EDR3 catalog provides a number of statistical tools that can be used to analyze the quality of the astrometric fit. Since the GAIA precision is sensitive to the brightness of the star, in Figure\,\ref{fig:gaia-stats}, we present key diagnostic metrics to evaluate the quality of the HR\,2562 EDR3 entry. Overall, the star appears well behaved, albeit with a small but significant astrometric excess noise of 0.158\,mas. It is not clear if this excess noise is astrophysical in nature, but the excess noise is significantly lower than reported in the {\it Hipparcos} catalogue (2.4\,mas). Overall, we consider the GAIA data to be of good quality, i.e., the data show no strong evidence for departure from a linear motion over the course of the GAIA observations.

\section{Full Orbital Fit}
\label{append:Full orbit fit}
We present the full corner plot to the \HRB orbit MCMC run in section \ref{subsec:coplanar}. 
\begin{figure}
    \centering
    \includegraphics[width=\textwidth]{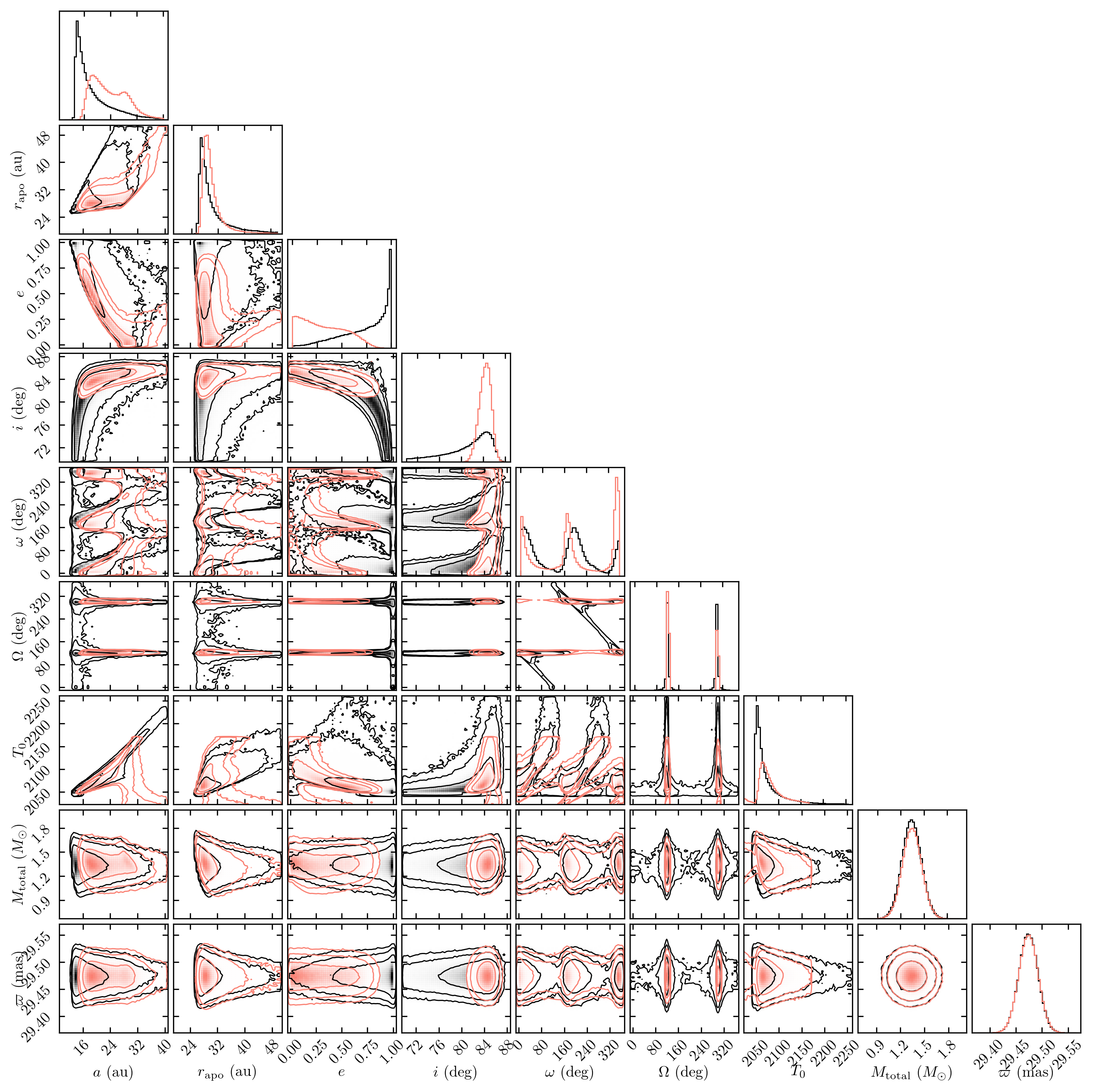}
    \caption{Full corner plot of HR2562 orbit with and without the new constraints from the ALMA fit.  Contours represent the 1, 2, and 3\,$\sigma$ confidence levels.}
    \label{fig:HR2562_corner_big}
\end{figure}

\bibliography{bibliography.bib}

\end{document}